\documentclass[aps,pre,floatfix]{revtex4}
\usepackage{epsf}
\usepackage{subfigure}
\usepackage{amsmath}
\usepackage{amssymb}
\usepackage{graphicx}
\usepackage{epstopdf}
\font\helvb=cmssbx12

\begin{document}

\newcommand{\be}{\begin{equation}}
\newcommand{\ee}{\end{equation}}
\newcommand{\bea}{\begin{eqnarray}}
\newcommand{\eea}{\end{eqnarray}}
\newcommand{\mean}[1]{\left \langle #1 \right \rangle}

\title{\bf Time-reversal symmetry relations for currents in quantum and stochastic nonequilibrium systems}

\author{Pierre Gaspard}
\affiliation{Center for Nonlinear Phenomena and Complex Systems \& Department of Physics,\\
Universit\'e Libre de Bruxelles, Code Postal 231, Campus Plaine, 
B-1050 Brussels, Belgium}

\begin{abstract}
An overview is given of recent advances in the nonequilibrium statistical mechanics of quantum systems and, especially, of time-reversal symmetry relations that have been discovered in this context. The systems considered are driven out of equilibrium by time-dependent forces or by coupling
to large reservoirs of particles and energy.  The symmetry relations are established for the exchange
of energy and particles between the subsystem and its environment.  These results have important
consequences. In particular, generalizations of the Kubo formula and the Casimir-Onsager
reciprocity relations can be deduced beyond linear response properties.
Applications to electron quantum transport in mesoscopic semiconducting circuits are discussed.
\end{abstract}

\noindent {R. Klages, W. Just, and C. Jarzynski, Eds.,\\ 
{\it Nonequilibrium Statistical Physics of Small Systems:} \\
{\it Fluctuation Relations and Beyond} \\
(Wiley-VCH, Weinheim, 2012; ISBN 978-3-527-41094-1)}

\vskip 0.5 cm

\maketitle

\section{Introduction}
\label{Intro}

Microreversibility is the symmetry under time reversal of the equations 
of motion for the microscopic particles composing matter.
This symmetry is a property of the electromagnetic interaction \cite{CPT}.  The motion is ruled by
Newton's equations for the positions and velocities of the particles in classical mechanics and by Schr\"odinger's equation for the wavefunction in quantum mechanics.  Classical mechanics emerges out of quantum mechanics in the limit where de Broglie's wavelength becomes smaller than the spatial scale of features in the energy landscape where each particle moves.  A particle of mass $m$ and velocity $v$ has its de Broglie wavelength given by
\be
\lambda = \frac{h}{mv} 
\ee
where $h$ is Planck's constant.  Hence, the wavelength is shorter for heavier particles in systems at higher temperature. At room temperature, the wavelength of nuclei is smaller than typical interatomic distances.  This is not the case for electrons which maintain the interatomic chemical bonds by the spatial extension of their quantum wavefunction.
Consequently, the motion of nuclei is essentially classical while electrons behave quantum mechanically.  At low temperature, de Broglie's wavelengths are larger.  For instance, in semiconducting nanodevices at subKelvin electronic temperatures, the nuclei are nearly frozen while the electronic waves propagate ballistically in artificial circuits of a few hundred nanometers \cite{I97,FGB09,NB09}.

Newton's and Schr\"odinger's equations have in common to be time-reversal symmetric and, moreover, deterministic in the sense that the time evolution they rule is uniquely determined by initial conditions belonging to either the classical phase space of positions and velocities, or the Hilbert space of wavefunctions.  In both schemes, the initial conditions themselves are not uniquely determined once for all, but free to take arbitrary values depending on the particular experiment considered.  In this regard, the equations of motion describe all the possible histories that a system may follow, among which there exists the unique history actually followed by the system under specific conditions.  Besides, most of the histories do not coincide with their time reversal in typical classical or quantum systems.  For instance, the trajectory of a free particle with a non-vanishing velocity is physically distinct from its time reversal in the phase space of classical states:
\be
({\bf r}_t,{\bf v}_t) =( {\bf v}_0 t + {\bf r}_0,{\bf v}_0)   \qquad\neq \qquad  ({\bf r}'_t,{\bf v}'_t) =(-{\bf v}_0 t + {\bf r}_0,-{\bf v}_0) \, .
\ee
If the first trajectory was coming from the star Antares towards the constellation of Taurus on the opposite side of the celestial sphere, the second would travel in the opposite direction following a reversed history \cite{G06}.  A similar result holds for the wavepacket of a free quantum particle.  For such systems, the selection of initial conditions breaks the time-reversal symmetry.  Indeed, the solutions of an equation may have a lower symmetry than the equation itself. This is well known in condensed matter physics as the phenomenon of spontaneous symmetry breaking.  For example in ferromagnetism, the orientation of magnetization is determined by fluctuations at the initial stage of the process, i.e., by the initial conditions.  

These considerations extend to statistical mechanics where the time evolution concerns statistical ensembles described, in classical mechanics, by a probability density $\rho$ evolving in phase space according to Liouville's equation:
\be
\partial_t\, \rho = \{ H,\rho\}_{\rm cl}
\ee
where $H$ is the Hamiltonian and $\{ \cdot\,,\cdot\}_{\rm cl}$ the Poisson bracket or, in quantum mechanics, by a density operator $\hat\rho$ evolving according to von Neumann's equation:
\be
\partial_t\, \hat\rho = \frac{1}{i\hbar} [ \hat H, \hat \rho ]
\label{vN_eq}
\ee
where $[\,\cdot\,,\cdot\,]$ is the commutator, $i=\sqrt{-1}$, and $\hbar=h/(2\pi)$ \cite{B75}.
Both equations are symmetric under time reversal but they may admit solutions that correspond to histories differing from their time reversals.

However, thermodynamic equilibria are described by Gibbsian statistical ensembles such as the canonical ensemble  defined by the probability distribution
\be
\rho = \frac{1}{Z} \, \exp(-\beta H) \qquad \mbox{with} \qquad \beta = \frac{1}{k_{\rm B} T}
\label{canonical}
\ee
for the temperature $T$ and where $k_{\rm B}$ is Boltzmann's constant.  This distribution is a stationary solution of Liouville's or von Neumann's equation and it is time-reversal symmetric.  As a consequence, the principle of detailed balancing holds according to which opposite fluctuations are equiprobable.  The entropy production vanishes and there is no energy dissipation on average in systems at equilibrium.  We notice that the thermodynamic equilibrium is a state which is stationary from a statistical viewpoint but dynamical from a mechanical viewpoint.

During the last decades, remarkable advances have been achieved in the understanding of the statistical properties of nonequilibrium systems. These results find their origins in the study of large-deviation properties of chaotic dynamical systems, for which methods have been developed to characterize randomness in time (instead of space or phase space as done in equilibrium statistical mechanics) \cite{ER85}.  On the basis of dynamical systems theory, relationships have been established between dynamical large-deviation quantities and transport properties such as diffusion, viscosity, and electric or heat conductivities \cite{PH88,GN90,GB95,DG95,GD95,vBDPD97,G98,D99,GCGD01,K07,EM08}.  In this context, the further consideration of time reversal led to the discovery of different types of symmetry relations depending on the nonequilibrium regime.  Indeed, systems may evolve out of equilibrium under different conditions:

\begin{itemize}

\item{} In isolated systems, the solution of Liouville's equation is not stationary and evolves in time if the initial distribution $\rho_0$ differs from an equilibrium one.  If the dynamics is mixing, the statistical distribution $\rho_t$  weakly converges towards a final equilibrium distribution after nonequilibrium transients.  Transient states with exponential decay exist in the forward or backward time evolutions, which are mapped onto each other by time reversal, while the equilibrium state is symmetric \cite{G98,D99,GCGD01}.

\item{} In systems controlled by time-dependent external forces, the statistical distribution $\rho_t$ remains out of equilibrium since the relaxation towards equilibrium is not possible.  For instance, the nonequilibrium work is a fluctuating quantity of interest in small systems such as single molecules subjected to the time-dependent forces of optical tweezers or atomic force microscopy \cite{CRJSTB05,LSBFLD11,J97,C98,C99,J00}.

\item{} In open systems in contact with several reservoirs at different temperatures or chemical potentials, a nonequilibrium steady state is reached after some relaxation time.  Such states are described by stationary statistical distributions that are no longer symmetric under time reversal contrary to the equilibrium canonical state (\ref{canonical}). Indeed, out of equilibrium, mean fluxes of energy or matter are flowing on average across the open system and this directionality breaks the time-reversal symmetry.  In small systems or on small scales, the energy or particle currents are fluctuating quantities described by the nonequilibrium statistical distribution.

\end{itemize}

Because the underlying microscopic dynamics is reversible, the fluctuations of nonequilibrium quantities obey remarkable relationships, which are valid not only close to equilibrium but also arbitrarily far from equilibrium \cite{J97,C98,C99,J00,ECM93,GC95,G96,K98,LS99,M99,ES02,vZC03,ADEN10,MMO11,J11}.  Several kinds of such relationships have been obtained for transitory or stationary nonequilibrium situations, for classical or quantum systems, and for Markovian or non-Markovian stochastic processes.

Moreover, further time-reversal symmetry relations have also been obtained for the probabilities of the histories or paths followed by a system under stroboscopic observations \cite{MN03,G04JSP,G05,KPV07,G07CRP,AGCGJP07,AGCGJP08,CRJMN04}.  It turns out that the time asymmetry of nonequilibrium statistical distributions implies that typical histories are more probable than their corresponding time reversal.  In this regard, dynamical order manifests itself away from equilibrium \cite{G07CRP}.  Furthermore, the breaking of detailed balancing between forward and reversed histories is directly related to the thermodynamic entropy production, as established for classical, stochastic and quantum systems \cite{MN03,G04JSP,G05,KPV07,G07CRP,AGCGJP07,AGCGJP08,CRJMN04}.

The purpose of the present chapter is to give an overview of the time-reversal symmetry relations established for nonequilibrium quantum systems \cite{CRJMN04,K00,T00,M03,TN05,EHM07,HEM07,SU08,AG08PRL,AGMT09,SD07,TH07,TLH07,CHT11,GK08,EHM09,GLSSISEDG06,GLSSIEDG09,FHTH06,UGMSFS10,KRBMGUIE12,NYHCKOLESUG10,NYHCKOLESUG11}.  Recently, important experimental and theoretical work has been devoted to transport in small quantum systems and, in particular, to electronic transport in quantum point contacts, quantum dots, and coherent quantum conductors \cite{I97,FGB09,NB09}.  In these systems, single-electron transfers can be observed experimentally and subjected to full counting statistics, allowing the experimental test of the time-reversal symmetry relations and their predictions.  Here, the aim is to provide a comprehensive overview of the results obtained till now on this topic and to discuss some of the open issues.

This chapter is organized as follows.  In Section~\ref{Fnal}, functional symmetry relations are established for quantum systems driven by time-dependent external forces in a constant magnetic field, which allows us to derive the Kubo formulae and the Casimir-Onsager reciprocity relations for the linear response properties.  In Section~\ref{trFT}, general methods are presented to describe open quantum systems in contact with several reservoirs and transitory fluctuation theorems are obtained.  In Section~\ref{stFT}, an appropriate long-time limit is taken in order to reach a nonequilibrium steady state.  In this limit, the stationary fluctuation theorem is obtained from the transitory one for all the currents flowing across the open system.  In Section~\ref{Resp}, the current fluctuation theorem is used to obtain the response properties.  The Casimir-Onsager reciprocity relations and fluctuation-dissipation relations are generalized from linear to nonlinear response properties.  In Section~\ref{Indep}, these results are obtained for independent electrons in quantum point contacts or quantum dots.  The generating function of full counting statistics is computed using Klich formula and the connection is established with the Levitov-Lesovik formula in the Landauer-B\"uttiker scattering approach.  The equivalence with the Keldysh approach is discussed.  In Section~\ref{Master}, the current fluctuation theorem is derived in the master-equation approach for the corresponding stochastic process.  The second law of thermodynamics is deduced from the stationary fluctuation theorem for the currents and results on the statistics of histories are presented. In Section~\ref{Electron}, the general theory is applied to electronic transport in quantum dots, quantum point contacts, and coherent quantum conductors.  Conclusions are drawn in Section~\ref{Conclusions}.

\section{Functional symmetry relations and response theory}
\label{Fnal}

We consider a quantum system described by the Hamiltonian operator $\hat H(t;\mathcal{B})$, which depends on the time $t$ and the external magnetic field $\mathcal{B}$. The time-reversal operator $\hat\Theta$ reverses the magnetic field but otherwise leaves invariant the Hamiltonian:
\be
\hat\Theta\, \hat H(t;\mathcal{B})\, \hat\Theta^{-1} = \hat H(t; -\mathcal{B}) \, .
\label{TR}
\ee
The reason is that the time-reversal symmetry of the electromagnetic interaction concerns the whole physical system including the external electric currents generating the magnetic field.  The magnetic field changes its sign because the external currents are reversed under time reversal.  At the level of the system itself, the symmetry should thus be implemented by the combination of the internal operator $\hat\Theta$ and the reversal of the background magnetic field.

Since the system is driven by time-dependent forces, a comparison should be carried out between some protocol, called the forward protocol, and the reversed protocol \cite{C98,C99,J00}.

In the {\it forward protocol}, the system starts in the equilibrium distribution:
\be
\hat\rho(0;\mathcal{B})=\frac{{\rm e}^{ -\beta \hat H(0;\mathcal{B})}}{Z(0)} 
\label{rho(0)}
\ee
at the inverse temperature $\beta = (k_{{\rm B}}T)^{-1}$ and the free energy $F(0)=-k_{\rm B}T \ln Z(0)$ with $Z(0)={\rm tr} \, {\rm e}^{ -\beta \hat H(0;\mathcal{B})}$.  The system evolves from the initial time $t=0$ until the final time $t={\cal T}$ under the unitary operator that is the solution of Schr\"odinger's equation:
\be
i\hbar \frac{\partial}{\partial t} \hat U_{\rm F} (t;\mathcal{B}) = \hat H(t;\mathcal{B}) \, \hat U_{\rm F} (t;\mathcal{B}) 
\label{U}
\ee
with the initial condition $\hat U_{\rm F}(0;\mathcal{B})=1$.
The average of an observable $\hat A$ is given by
\be
\mean{\hat A_{\rm F}(t)}= {\rm tr} \, \hat\rho(0) \, \hat A_{\rm F}(t) 
\ee
where
\be
\hat A_{\rm F}(t)=\hat U^{\dagger}_{\rm F}(t) \, \hat A\,  \hat U_{\rm F}(t)
\ee
is the operator in the Heisenberg representation.

In the {\it reversed protocol}, the system starts in the other equilibrium distribution:
\be
\hat\rho({\cal T};-\mathcal{B})=\frac{{\rm e}^{ -\beta \hat H({\cal T};-\mathcal{B})}}{Z({\cal T})}
\label{rho(T)}
\ee
of free energy $F({\cal T})=-k_{\rm B}T \ln Z({\cal T})$ with $Z({\cal T})={\rm tr} \, {\rm e}^{ -\beta \hat H({\cal T};-\mathcal{B})}$ and evolves under the reversed time evolution operator obeying 
\be
i\hbar \frac{\partial}{\partial t} \hat U_{\rm R} (t;\mathcal{B}) = \hat H({\cal T}-t;\mathcal{B})\, \hat U_{\rm 
R} (t;\mathcal{B}) 
\label{UR}
\ee
with the initial condition $\hat U_{\rm R} (0;\mathcal{B})=1$. In the reversed protocol,
the system thus follows the reversed external driving in the reversed magnetic field, 
starting at the time $t=0$ with the Hamiltonian $\hat H({\cal T};-\mathcal{B})$
and ending at the time $t={\cal T}$ with the Hamiltonian $\hat H(0;-\mathcal{B})$.

We have the\\

{\bf Lemma:} \cite{AG08PRL}
{\it The forward and reversed time evolution unitary operators are related to 
each other according to}
\be
\hat U_{\rm R}(t;-\mathcal{B}) = \hat\Theta \, \hat U_{\rm F} ({\cal T}-t;\mathcal{B}) \, \hat U^{\dagger}_{\rm F} ({\cal T};\mathcal{B}) \, \hat \Theta^{-1} 
\qquad\mbox{\it with} \qquad 0 \leq t \leq {\cal T} \; .
\label{lUUl}
\ee

This lemma is proved by using the antiunitarity of the time-reversal operator, Eqs.~(\ref{U}) and (\ref{UR}) obeyed by the unitary evolution operators, as well as their initial conditions \cite{AG08PRL}.

Let us consider a time-independent observable $\hat A$ with a definite parity under
time reversal, i.e., such that $\hat\Theta \, \hat A \, \hat\Theta^{-1} = \epsilon_A  \,\hat A$ with $\epsilon_A = \pm 1$.
The lemma allows us to relate its forward and reversed Heisenberg representations according to
\be
\hat A_{\rm F}(t)= \epsilon_A \,
\hat U^{\dagger}_{\rm F}({\cal T})\, \hat \Theta^{-1} \, \hat A_{\rm R}({\cal T}-t) \, \hat\Theta \, 
\hat U_{\rm F}({\cal T}) \, .
\label{Ar}
\ee
In this setting, we have the\\

{\bf Theorem:} \cite{AG08PRL}
{\it If $\lambda(t)$ denotes an arbitrary function of time, the following functional relation holds:}
\be
\mean{{\rm e}^{\int_0^{\cal T} dt\,  \lambda(t)\, \hat A_{\rm F}(t)} \, {\rm e}^{- \beta 
\hat H_{\rm F}({\cal T})} \, {\rm e}^{ \beta \hat H(0)}}_{\rm F, \mathcal{B}} 
= {\rm e}^{-\beta \Delta F}
\mean{ {\rm e}^{\epsilon_A \int_0^{\cal T} dt\, \lambda({\cal T}-t)\, \hat A_{\rm R}(t)} 
}_{\rm R, -\mathcal{B}} 
\label{qw}
\ee
{\it where $\hat H_{\rm F}({\cal T})=\hat U^{\dagger}_{\rm F}({\cal T}) \,\hat H({\cal T};\mathcal{B}) \,\hat U_{\rm F}({\cal T})$ and $\Delta F=F({\cal T})-F(0)$ is the free-energy difference between the initial equilibrium states (\ref{rho(T)}) and (\ref{rho(0)}) of the reversed and forward protocols.}
\\

This theorem, which has been demonstrated in Ref.~\cite{AG08PRL}, 
extends results previously obtained in the restricted case 
where there is no change in free energy $\Delta F=0$ \cite{BK77,S94}.

Moreover, the quantum Jarzynski equality is deduced in the special case 
where $\lambda=0$ in Eq.~(\ref{qw}):
\be
\mean{{\rm e}^{- \beta \hat H_{\rm F}({\cal T})} \, {\rm e}^{ \beta \hat H(0)}}_{\rm F, \mathcal{B}} 
= {\rm e}^{-\beta \Delta F} \, .
\label{QJ}
\ee
In the left-hand side, the quantity inside the bracket can be interpreted in terms of the work performed on the system in the quantum scheme where von Neumann measurements of the energy are carried out at the initial and final times \cite{K00,T00,M03,TH07,TLH07,CHT11,GK08}.
In the classical limit where the operators commute, both exponential functions combine into the exponential of the classical work $W_{\rm 
cl}=\left[H_{\rm F}({\cal T})-H(0)\right]_{\rm cl}$ and the classical Jarzynski equality \cite{J97}
\be
\mean{{\rm e}^{- \beta W_{\rm cl}}}_{\rm F, \mathcal{B}} 
= {\rm e}^{-\beta \Delta F} 
\ee
 is recovered for the nonequilibrium work performed on the system during the forward protocol.  We notice that $\Delta F$ does not depend on the sign of the magnetic field because the canonical equilibrium distributions are time-reversal symmetric.
 
 Remarkably, the functional symmetry relation (\ref{qw}) unifies the work relations and the response theory in a common framework. Indeed, the Kubo formulae as well as the Casimir-Onsager reciprocity relations can also be deduced from this relation.  With this aim, we assume that the system is composed of $N$ particles of electric charges $\{e_n\}_{n=1}^N$ subjected to an external time-dependent electric field ${\cal E}_{\nu}(t)$ in the spatial direction $\nu=1,2$ or $3$. The electric field is supposed to vanish at the initial and final times ${\cal E}_{\nu}(0)={\cal E}_{\nu}({\cal T})=0$ so that the free-energy difference is here equal to zero, $\Delta F=0$.  The time-dependent Hamiltonian is given by
 \be
 \hat H(t) = \hat H_0 - {\cal E}_{\nu}(t) \sum_{n=1}^N e_n \, \hat x_{n\nu}
 \ee
 with the position operators $\hat x_{n\nu}$ of the $N$ particles.  The observable we consider is the electric current density in the spatial direction $\mu=1,2$ or $3$:
 \be
 \hat A = \frac{1}{V} \, \hat J_{\mu} = \frac{1}{V} \, \sum_{n=1}^N e_n \, \frac{d\hat x_{n\mu}}{dt}
 \ee
 where $V$ is the volume of the system.
In order to obtain the linear response of the observable $\hat A$ with 
respect to the perturbation due to an electric field
${\cal E}_{\nu}(t)$ of small amplitude, the functional derivative of Eq. (\ref{qw}) is taken with respect to $\lambda({\cal T})$ around $\lambda=0$, which yields
\be
\mean{\hat A_{\rm F}({\cal T}) \, {\rm e}^{- \beta \hat H_{\rm F}({\cal T})}\, {\rm e}^{ \beta 
\hat H_0}  }_{\rm F, \mathcal{B}}
= \mean{\hat A}_{{\rm eq}, \mathcal{B}} \, .
\label{Alin}
\ee
Developing this expression to first order in the electric field, the mean value of the current density at the final time $t={\cal T}$ is obtained as \cite{AG08PRL}
\be
j_{\mu}({\cal T})=\mean{\hat A_{\rm F}({\cal T})}_\mathcal{B} = \mean{\hat A}_{{\rm eq},\mathcal{B}}+ \int_0^{\cal T} dt \; {\cal E}_{\nu}({\cal T}-t)\; \phi_{\mu\nu}(t;\mathcal{B}) + O({\cal E}_{\nu}^2)
\label{greenab}
\ee
with the response function
\bea
\phi_{\mu\nu}(t;\mathcal{B})=\frac{1}{V}\, \int_0^\beta du \ \langle \hat J_{\nu}(-i\hbar 
u)\, \hat J_{\mu}(t)\rangle_{{\rm eq},\mathcal{B}}  \, .
\label{phi}
\eea
Since the mean value of the current is vanishing at equilibrium $\mean{\hat A}_{{\rm eq}, \mathcal{B}}=0$, we find the expression of Ohm's law
\be
j_{\mu} = \sigma_{\mu\nu} \, {\cal E}_{\nu}
\ee
in the long-time limit ${\cal T}\to\infty$.  The electric conductivity is given by
\be
\sigma_{\mu\nu}=\int_0^{\infty}\phi_{\mu\nu}(t;\mathcal{B}) \, dt =\sigma_{\mu\nu}(\omega=0;\mathcal{B})
\ee
in terms of Kubo's formulae for the ac~conductivities \cite{K57}
\be
\sigma_{\mu\nu}(\omega;\mathcal{B})  = \frac{1}{V}\, \int_0^{\infty} dt \, {\rm e}^{i\omega t} \, \int_0^\beta du \ \langle \hat J_{\nu}(-i\hbar 
u)\, \hat J_{\mu}(t)\rangle_{{\rm eq},\mathcal{B}} \; .
\label{sigma}
\ee
These conductivities satisfy  the Casimir-Onsager reciprocity relations \cite{O31,C45} 
\be
\sigma_{\mu\nu}(\omega;\mathcal{B})=\sigma_{\nu\mu}(\omega;-\mathcal{B}) \, .
\ee
In this way, the linear response properties are recovered 
from the functional symmetry relation (\ref{qw}).
Higher-order terms in the expansion can be obtained as well for the nonlinear response properties.  

The functional time-reversal symmetry relation thus provides a
unifying framework to study the fundamental properties of quantum systems driven out of equilibrium by time-dependent external fields.

\section{The transitory current fluctuation theorem}
\label{trFT}

\begin{figure}[h]
\begin{center}
\includegraphics[scale=0.45]{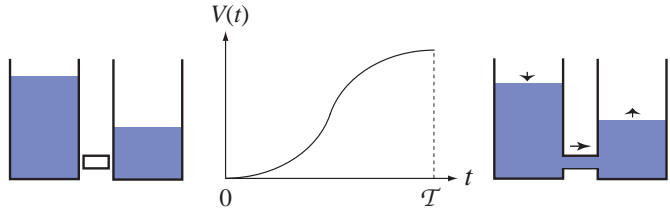}
\caption{Schematic representation of the coupling of a subsystem to a pair of reservoirs by the potential $\hat V(t)$ during the time interval $0\leq t \leq {\cal T}$, for the protocol used to deduce the {\it transitory} current fluctuation theorem.  At the initial time $t=0$, the potential is vanishing, $\hat V(0)=0$. At the final time $t={\cal T}$, the potential may be non vanishing, but then given by a sum of operators separately acting on the quantum state spaces of the subsystem and the reservoirs.}
\label{fig1}
\end{center}
\end{figure}

A related problem of interest is the transport of particles and energy across
an open quantum subsystem in contact with several reservoirs at different temperatures and chemical potentials.  The subsystem is coupled to the reservoirs by a time-dependent potential $\hat V(t)$ during some time interval $0\leq t \leq {\cal T}$.  The total system evolves in time under the quantum-mechanical unitary time evolution from the initial state specified by taking an equilibrium distribution in each separate part of the total system until the final time $\cal T$, as schematically depicted in Fig.~\ref{fig1}.  Here, the general case is considered where the time-dependent potential may leave some permanent changes inside the subsystem and the reservoirs after the final time $t={\cal T}$.
The flow of energy and particles between the reservoirs is determined by two successive von Neumann quantum measurements in each part of the total system, the first at the initial time $t=0$ and the second at the final time $\cal T$.  In this setup, a time-reversal symmetry relation is established for the transfers of energy and particles between the parts.

The total Hamiltonian operator is given by
\be
\hat H(t;{\cal B}) =
\left\{
\begin{array}{lcc}
\hat{\cal H}_{\rm S} + \sum_{j=1}^r \hat{\cal H}_j &\mbox{for} & t \leq 0 \, ,\\
\hat{\cal H}_{\rm S} + \sum_{j=1}^r \hat{\cal H}_j + \hat V(t) & \mbox{for} & 0 < t < {\cal T} \, ,\\
\hat{\cal H}'_{\rm S} + \sum_{j=1}^r \hat{\cal H}'_j & \mbox{for} & {\cal T} \leq t \, ,
\end{array}
\right.
\label{H_tot}
\ee
where $\hat{\cal H}_{\rm S}$ and $\hat{\cal H}_j$  denote the Hamiltonian operators of the isolated subsystem and the $j^{\rm th}$ reservoir before the interaction is switched on. $\hat V(t)$ is the time-dependent interaction between the subsystem and the reservoirs.  After the final time $t={\cal T}$, the subsystem and the reservoirs no longer interact with each other, but the final potential $\hat V({\cal T})$ may have added possible contributions to the Hamiltonian operators of the subsystem and the reservoirs, $\hat{\cal H}'_{\rm S}$ and $\hat{\cal H}'_j$, which could thus differ from the initial ones.  Moreover, the total Hamiltonian always obeys the time-reversal symmetry (\ref{TR}).

The observables of the total system include not only the Hamiltonian operators
but also the particle numbers of species $\alpha=1,2,...,c$
inside the subsystem, $\hat{\mathcal N}_{{\rm S}\alpha}$, and the reservoirs,
$\hat{\mathcal N}_{j\alpha}$ with $j=1,2,...,r$.  These latter observables are also time-reversal symmetric: $\hat\Theta\, \hat{\mathcal N}_{j\alpha} \, \hat\Theta^{-1} = \hat{\mathcal N}_{j\alpha}$ with $j={\rm S},1,2,...,r$.

The total numbers of particles are given by
\be
\hat N_{\alpha} = \hat{\mathcal N}_{{\rm S}\alpha} + \sum_{j=1}^r \hat{\mathcal N}_{j \alpha}
\qquad \mbox{for}\qquad \alpha=1,2,...,c
\ee
and they commute with the total Hamiltonian operator because they are conserved:
\be
[ \hat H(t;{\cal B}), \hat N_{\alpha} ] = 0 \; .
\ee
The system may admit further constant numbers of particles if the reservoirs are not all connected to each other.  If the total system was composed of $d$ disconnected parts between which there is no flux of particles, the particle numbers of every species would be separately conserved in each part so that the system would admit $c\times d$ independent constants of motion.

Moreover, before the initial time and after the final time, the reservoirs are decoupled from each other so that their Hamiltonian operator commutes with the particle numbers:
\be
\left[ \hat{\mathcal H}_j, \hat{\mathcal N}_{j'\alpha} \right] = 0 \quad \mbox{for}\quad t<0 \qquad\qquad  \mbox{and}\qquad\qquad  \left[ 
\hat{\mathcal H}'_j, \hat{\mathcal N}_{j'\alpha} \right] = 0 
\quad \mbox{for} \quad {\cal T}<t
\ee
for every $j,j'={\rm S},1,2,...,r$ and $\alpha=1,2,...,c$.

In nonequilibrium systems, the currents are determined by the differences
of temperature and chemical potentials between the reservoirs.
Since the reservoirs are large, the initial temperature
and chemical potentials of the subsystem 
are not relevant if the subsystem is small enough.
In this regard, we can simplify the formulation of the problem 
by regrouping the subsystem with one of the reservoirs, for instance the first one, and
redefine the Hamiltonian and particle-number operators as follows:
\be
\begin{array}{llll}
\hat H_1 = \hat{\mathcal H}_{\rm S} + \hat{\mathcal H}_1 \, ,\qquad \qquad & \hat H'_1 = \hat{\mathcal H}'_{\rm S} + \hat{\mathcal H}'_1 \, ,\qquad \qquad & \hat N_{1\alpha} = \hat{\mathcal N}_{{\rm S}\alpha} + \hat{\mathcal N}_{1\alpha} \, ,&  \qquad\qquad \mbox{for} \quad  j=1\qquad\qquad\mbox{and} \\
\hat H_j  = \hat{\mathcal H}_j \, , \qquad \qquad & \hat H'_j  = \hat{\mathcal H}'_j \, , \qquad \qquad & \hat N_{j\alpha} =  \hat{\mathcal N}_{j\alpha} \, ,& \qquad\qquad \mbox{for} \quad  j=2,...,r \, . 
\end{array}
\ee
In this case, the total Hamiltonian given by Eq. (\ref{H_tot}) can be rewritten as
\be
\hat H(t;{\cal B}) =
\left\{
\begin{array}{lcc}
\sum_{j=1}^r \hat H_j &\mbox{for} & t \leq 0 \, ,\\
\sum_{j=1}^r \hat H_j + \hat V(t) & \mbox{for} & 0 < t < {\cal T} \, ,\\
\sum_{j=1}^r \hat H'_j & \mbox{for} & {\cal T} \leq t \, ,
\end{array}
\right.
\label{H_tot_bis}
\ee
and the total particle numbers of species $\alpha=1,2,...,c$ as
\be
\hat N_{\alpha} = \sum_{j=1}^r \hat N_{j \alpha} \, .
\label{N_a,tot}
\ee

As in previous Section~\ref{Fnal}, a comparison is made between the outcomes of some forward and  reversed protocols.

The {\it forward protocol} starts with the total system in the grand-canonical equilibrium state of the decoupled parts at the different inverse temperatures $\beta_j=(k_{\rm B}T_j)^{-1}$ and chemical potentials $\mu_{j\alpha}$: 
\be
\hat\rho(0;{\cal B}) = \prod_{j} \Xi_j({\cal B})^{-1} \, {\rm 
e}^{-\beta_j\left[\hat H_j({\cal B})-\sum_{\alpha}\mu_{j\alpha} \hat N_{j\alpha}\right]}
= \prod_{j} {\rm e}^{-\beta_j\left[\hat H_j({\cal B})-\sum_{\alpha}\mu_{j\alpha} 
\hat N_{j\alpha}-\Phi_j({\cal B})\right]}
\label{rho_0}
\ee
where $\Phi_j({\cal B})= -k_{\rm B}T_j \ln \Xi_j({\cal B})$ denotes the grand-canonical
thermodynamic potential of the $j^{\rm th}$ part in the 
initial equilibrium state.  We notice that the grand-canonical potential is even in the magnetic field if the corresponding Hamiltonian has the time-reversal symmetry (\ref{TR}).  The system evolves from the initial time $t=0$ until the final time $t={\cal T}$ under the forward time evolution unitary operator obeying Eq.~(\ref{U}) with the initial condition $\hat U_{\rm F}(0;\mathcal{B})=1$.

In order to determine the fluxes of energy and particles,
quantum measurements are carried out on the reservoirs
before and after the unitary time evolution.

An {\it initial quantum measurement} is performed which prepares the system 
in the eigenstate $\vert\Psi_k\rangle$ of the energy and particle-number operators:
\bea
t \leq 0: \qquad  \hat H_j \vert\Psi_k\rangle &=& \varepsilon_{jk} \vert\Psi_k\rangle \; ,\\
\hat N_{j\alpha} \vert\Psi_k\rangle &=& \nu_{j\alpha k} \vert\Psi_k\rangle \; .
\eea

After the time interval $0 < t <{\cal T}$, a {\it final quantum measurement} is performed in which the system is observed in the eigenstate $\vert\Psi'_l\rangle$ of the final energy and particle-number operators:
\bea
{\cal T}\leq t: \qquad   \hat H'_j \vert\Psi'_l\rangle &=& 
\varepsilon'_{jl} \vert\Psi'_l\rangle \; ,\\
\hat N_{j\alpha} \vert\Psi'_l\rangle &=& \nu'_{j\alpha l} 
\vert\Psi'_l\rangle \; .
\eea
We notice that semi-infinite time intervals are available to perform
the initial and final quantum measurements of
well-defined eigenvalues.  Accordingly, this scheme based 
on two quantum measurements provides a systematic way to measure 
the energies and the numbers of particles transferred between the reservoirs
during the time interval $0 < t <{\cal T}$ of their mutual interaction.
During the forward protocol, the energy and the particle numbers are observed to vary in the $j^{\rm th}$ part of the system by the amounts
\bea
&&\Delta \varepsilon_j = \varepsilon'_{jl} - \varepsilon_{jk} \; ,
\label{De} \\
&&\Delta \nu_{j\alpha} = \nu'_{j\alpha l} - \nu_{j\alpha k} \; .
\label{Dn}
\eea
The probability distribution function to observe these variations is defined as
\bea
P_{\rm F} (\Delta \varepsilon_j,\Delta \nu_{j\alpha};{\cal B}) &\equiv& 
\sum_{kl} \; \prod_j \delta\left[\Delta\varepsilon_j - 
(\varepsilon'_{jl} - \varepsilon_{jk})\right] \; \prod_{j\alpha} 
\delta\left[ \Delta \nu_{j\alpha} - (\nu'_{j\alpha l} - 
\nu_{j\alpha k})\right] \nonumber\\ && \times \vert \langle 
\Psi'_l({\cal B}) \vert \hat U_{\rm F}({\cal T};{\cal B})\vert\Psi_k({\cal B})\rangle\vert^2 \; 
\langle \Psi_k({\cal B})\vert \hat \rho(0;{\cal B})\vert\Psi_k({\cal B})\rangle
\label{p_F}
\eea
in terms of Dirac distributions $\delta(\cdot)$.

In the {\it reversed protocol}, the system starts in the other equilibrium distribution:
\be
\hat \rho({\cal T};-{\cal B}) = \prod_{j} \Xi'_j({\cal B})^{-1}\, {\rm e}^{-\beta_j\left[\hat H'_j(-{\cal B})-\sum_{\alpha}\mu_{j\alpha} \hat N_{j\alpha}\right]}= 
\prod_{j} {\rm e}^{-\beta_j\left[\hat H'_j(-{\cal B})-\sum_{\alpha}\mu_{j\alpha} \hat N_{j\alpha}-\Phi'_j({\cal B})\right]}
\label{rho_T}
\ee
at the same inverse temperatures $\beta_j=(k_{\rm B}T_j)^{-1}$ and 
chemical potentials $\mu_{j\alpha}$ as in the forward protocol.  Here, 
$\Phi'_j({\cal B})= -k_{\rm B}T_j \ln \Xi'_j({\cal B})$ denotes 
the grand-canonical thermodynamic potential of the $j^{\rm th}$ 
part in the final equilibrium state and reversed magnetic field.
The reversed time evolution unitary operator obeys Eq.~(\ref{UR})
and it is related to the one of the forward protocol by Eq.~(\ref{lUUl}) with $t={\cal T}$ of the Lemma in Section~\ref{Fnal}.  Before and after the reversed protocol, quantum measurements are similarly carried out in order to determine the variations of energy and particle numbers in the reservoirs.  The probability distribution function of these variations during the reversed protocol is defined by an expression similar to Eq.~(\ref{p_F}).

Comparing the probability distribution functions of opposite variations during the forward and reversed protocols, the following symmetry relation is obtained
\be
\frac{P_{\rm F} (\Delta \varepsilon_j,\Delta \nu_{j\alpha};{\cal B})}
{P_{\rm R} (-\Delta \varepsilon_j,-\Delta \nu_{j\alpha};-{\cal B})}
 ={\rm e}^{\sum_j\beta_j(\Delta\varepsilon_{j}-\sum_{\alpha}\mu_{j\alpha}\Delta\nu_{j\alpha}-\Delta\Phi_{j})} 
\label{PFT}
\ee
with the differences $\Delta\Phi_{j} \equiv \Phi'_j- \Phi_j$ in the grand potentials \cite{AGMT09,EHM09}.

For the forward and reversed protocols, the {\it generating functions of the statistical moments} are defined as
\be
{\cal G}_{\rm F,R}(\xi_j,\eta_{j\alpha};{\cal B}) \equiv \int \prod_{j\alpha} 
d\Delta\varepsilon_j \, d\Delta \nu_{j\alpha} \, {\rm e}^{-\sum_j\xi_j 
\Delta\varepsilon_j -\sum_{\alpha} \eta_{j\alpha}\Delta \nu_{j\alpha}} \,
P_{\rm F,R} (\Delta \varepsilon_j,\Delta \nu_{j\alpha};{\cal B})
\label{G_FR}
\ee
in terms of the so-called {\it counting parameters} $\{\xi_j,\eta_{j\alpha}\}$.
The statistical moments of the energy and particle-number variations can be obtained
from this generating function by taking derivatives with respect to these counting parameters.  Now, the symmetry relation (\ref{PFT}) can be equivalently expressed as
\be
{\cal G}_{\rm F}(\xi_j,\eta_{j\alpha};{\cal B}) = {\rm 
e}^{-\sum_j\beta_j\Delta\Phi_{j}} \, {\cal G}_{\rm 
R}(\beta_j-\xi_j,-\beta_j\mu_{j\alpha}-\eta_{j\alpha};-{\cal B}) 
\label{GFT}
\ee
in terms of the temperatures and chemical potentials of the reservoirs.
We point out that this relation is only symmetric with respect to the inverse temperatures and the chemical potentials of the reservoirs.  

Such symmetry relations are useful for quantum system driven by time-dependent external forces.  If the relation (\ref{PFT}) is restricted to the energy variation, 
the fluctuating quantity is the work $W$ performed on the 
system and we recover the quantum version of Crooks' fluctuation 
theorem \cite{C99}
\be
\frac{P_{\rm F} (W;{\cal B})}{P_{\rm R} (-W;-{\cal B})} ={\rm e}^{\beta(W-\Delta F)}
\label{CrooksFT}
\ee
with the free-energy difference $\Delta F=F'-F$.  We notice that this relation implies Eq.~(\ref{QJ}) \cite{CHT11}. Similar symmetry relations may be obtained for cold atoms or molecules in rotating frames,
in which case the rotation rate $\Omega$ plays the role of the magnetic field $\cal B$.

\section{From the transitory to the stationary current fluctuation theorem}
\label{stFT}

If $r$ infinitely large reservoirs are coupled together via the subsystem of interest, a stationary state will be reached in the long-time limit \cite{T01,TM03,TT06}.  According to thermodynamics, the whole system is at equilibrium if the temperature as well as the chemical potentials of the different particle species are all uniform, i.e., if every reservoir shares the same temperature and chemical potentials.  This is no longer the case if the reservoirs have different temperatures or chemical potentials whereupon energy and particles are exchanged between the reservoirs across the subsystem coupling them together.  Therefore, energy and particle currents are induced by the so-called thermodynamic forces or affinities defined in terms of the differences of temperatures and chemical potentials with respect to some reference values \cite{DD36,P67,C85}.

\begin{figure}[h]
\begin{center}
\includegraphics[scale=0.45]{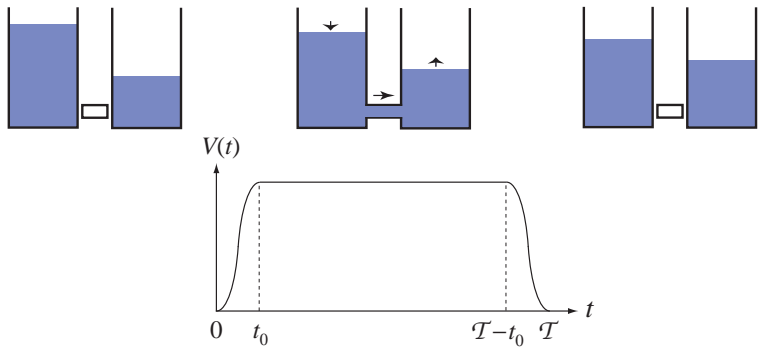}
\caption{Schematic representation of the coupling of a subsystem to a pair of reservoirs by the potential $\hat V(t)$ during the time interval $0\leq t \leq {\cal T}$, for the protocol used to obtain the {\it stationary} current fluctuation theorem.  Here, the time-dependent potential has the symmetry $\hat V(t)=\hat V({\cal T}-t)$, remaining constant during the time interval $t_0\leq t \leq {\cal T}-t_0$, and vanishing at the initial and final times: $\hat V(0)=\hat V({\cal T})=0$.}
\label{fig2}
\end{center}
\end{figure}

If we consider an open subsystem in contact with $r=2$ reservoirs as depicted in Fig.~\ref{fig2}, the {\it thermodynamic forces} or {\it affinities} are defined as
\bea
&& A_E \equiv \beta_1 - \beta_2 \; , \label{th-aff}\\
&& A_{\alpha} \equiv \beta_2\mu_{2\alpha} - \beta_1\mu_{1\alpha} \qquad \mbox{for}\qquad \alpha=1,2,...,c \; , \label{chem-aff}
\eea
driving respectively the energy and particle currents from reservoir No.~$2$ to reservoir No.~$1$.
The nonequilibrium conditions are thus specified by these affinities.

At the microscopic level of description, this system is described by the following Hamiltonian operator:
\be
\hat H(t;{\cal B}) = \hat H_1 + \hat H_2 + \hat V(t) \qquad\mbox{for}\qquad 0 \leq t \leq{\cal T} \; .
\ee
Here, the interaction $\hat V(t)$ is switched on during a short time interval $0\leq t<t_0$ at the beginning of the forward protocol with $t_0\ll{\cal T}$, remains constant $\hat V(t)=\hat V_0$ during most of the protocol for $t_0\leq t \leq {\cal T}-t_0$, and is switched off at the end when ${\cal T}-t_0 < t\leq {\cal T}$, as shown in Fig.~\ref{fig2}.  Moreover, the time dependence of the interaction is assumed to be symmetric under the transformation $t\to {\cal T}-t$.  Therefore, the forward and reversed protocols are identical except for the reversal of the magnetic field.  Consequently, $\Delta\Phi_j=0$ and the functions (\ref{G_FR}) of the forward and reversed protocols coincide, ${\cal G}_{\rm F}={\cal G}_{\rm R}\equiv {\cal G}_{\cal T}$.  In this case, the symmetry relation (\ref{GFT}) becomes
\be
{\cal G}_{\cal T}(\xi_1,\xi_2,\eta_{1\alpha},\eta_{2\alpha};{\cal B}) = {\cal G}_{\cal T}(\beta_1-\xi_1,\beta_2-\xi_2,-\beta_1\mu_{1\alpha}-\eta_{1\alpha},-\beta_2\mu_{2\alpha}-\eta_{2\alpha};-{\cal B}) \; . 
\label{GFT-st}
\ee
For applying to nonequilibrium steady states, the shortcoming is that this relation is not a symmetry with respect to the affinities or thermodynamic forces driving the system out of equilibrium. Accordingly, this relation does not yet concern nonequilibrium steady states associated with given affinities and further considerations are required.

We introduce the {\it cumulant generating function} as
\be
{\cal Q} \equiv \lim_{{\cal T}\to\infty} \, -\frac{1}{\cal T} \, \ln \, {\cal G}_{\cal T} \; .
\label{QC-dfn}
\ee
According to Eq.~(\ref{G_FR}), the statistical averages of the energy and particle-number variations are given by
\be
\frac{\partial{\cal Q}}{\partial\xi_j}\Big\vert_0 = \lim_{{\cal T}\to\infty} \frac{\langle \Delta\varepsilon_j\rangle}{\cal T}  \qquad\mbox{and}\qquad
\frac{\partial{\cal Q}}{\partial\eta_{j\alpha}}\Big\vert_0 = \lim_{{\cal T}\to\infty} \frac{\langle \Delta\nu_{j\alpha}\rangle}{\cal T}
\label{Aver}
\ee
while further differentiations give the cumulants. Besides, the short time interval $t_0$ is assumed to be constant in the long-time limit ${\cal T}\to\infty$.
Remarkably, it is possible to prove that
\\

{\bf Proposition:} \cite{AGMT09}  {\it If the long-time limit (\ref{QC-dfn}) exists, the cumulant generating function only depends on the differences $\xi=\xi_1-\xi_2$ and $\eta_{\alpha}=\eta_{1\alpha}-\eta_{2\alpha}$ between the counting parameters associated with the reservoirs:}
\be
{\cal Q}(\xi,\eta_{\alpha};{\cal B})  =  
 \lim_{{\cal T}\to\infty} \, -\frac{1}{\cal T} \, \ln \, {\cal G}_{\cal T}\left(\xi_0+\frac{\xi}{2},\xi_0-\frac{\xi}{2},\eta_{0\alpha}+\frac{\eta_{\alpha}}{2},\eta_{0\alpha}-\frac{\eta_{\alpha}}{2};{\cal B}\right)
 \qquad\mbox{\it for any} \quad \{\xi_0,\eta_{0\alpha}\}
\label{Q-G_tilde} 
\ee
{\it and the cumulant generating function of the energy and particle currents is given by}
\be
{\cal Q}(\xi,\eta_{\alpha};{\cal B}) = \lim_{t\to\infty} \, -\frac{1}{t} \, \ln \, \tilde {\cal G}_t(\xi,\eta_{\alpha};{\cal B})
\label{Q-dfn}
\ee
{\it where}
\be
\tilde {\cal G}_t(\xi,\eta_{\alpha};{\cal B}) = {\rm tr}\, \hat\rho_0 \, {\rm e}^{i\hat H t} \, {\rm e}^{-\frac{\xi}{2}(\hat H_1-\hat H_2)-\sum_{\alpha}\frac{\eta_{\alpha}}{2}(\hat N_{1\alpha}-\hat N_{2\alpha})} \, {\rm e}^{-i\hat H t} \, {\rm e}^{\frac{\xi}{2}(\hat H_1-\hat H_2)+\sum_{\alpha}\frac{\eta_{\alpha}}{2}(\hat N_{1\alpha}-\hat N_{2\alpha})}
\label{G_tilde}
\ee
{\it with $t={\cal T}-2t_0$, $\hat H=\hat H_1+\hat H_2 + \hat V_0$, and $\hat\rho_0$ is the initial grand-canonical distribution (\ref{rho_0}), which fixes the temperatures and the chemical potentials of the reservoirs.}
\\

This proposition is proved by bounding together the functions (\ref{GFT-st}) and (\ref{G_tilde}) according to
\be
L \, \tilde {\cal G}_t \leq {\cal G}_{\cal T} \leq K \, \tilde {\cal G}_t 
\label{inequal}
\ee
where $\tilde {\cal G}_t=\tilde {\cal G}_t(\xi_1-\xi_2,\eta_{1\alpha}-\eta_{2\alpha};{\cal B})$ and the two factors $L$ and $K$ are independent of the time interval $\cal T$ \cite{AGMT09}.  Therefore, the contributions of the factors $L$ and $K$ disappear in the long-time limit ${\cal T}\to\infty$.
Moreover, $\lim_{{\cal T}\to\infty}(t/{\cal T})=1$ since $t={\cal T}-2t_0$ and $t_0$ is constant.  For these reasons, the results (\ref{Q-G_tilde}) and (\ref{Q-dfn}) are obtained. Q.~E.~D.  

The inequalities (\ref{inequal}) are established in several steps \cite{AGMT09}.  The first two steps consist in removing the contributions from the initial and final lapses of time $t_0$, during which the interaction is switched on or off.  Indeed, these contributions become negligible in the long-time limit.  The third step has the effect of letting appear the differences $\xi_1-\xi_2$ and $\eta_{1\alpha}-\eta_{2\alpha}$ between the counting parameters of the two reservoirs, which is the main result of the proposition.  Accordingly, Eqs.~(\ref{Q-dfn}) and (\ref{G_tilde}) genuinely concern the fluxes of energy and particles transferred between the reservoirs.

Now, the symmetry relation (\ref{GFT-st}) implies the following
\\

{\bf Current fluctuation theorem:} \cite{AGMT09}  {\it As the consequence of the time-reversal symmetry $\hat\Theta\, \hat H({\cal B}) \, \hat\Theta^{-1}=\hat H(-{\cal B})$, the cumulant generating function (\ref{Q-dfn}) of the energy and particle currents satisfies the symmetry relation:}
\be
{\cal Q}(\xi,\eta_{\alpha};{\cal B}) = {\cal Q}(A_E-\xi,A_{\alpha}-\eta_{\alpha};-{\cal B})
\label{CFT}
\ee
{\it with respect to the thermal and chemical affinities (\ref{th-aff}) and (\ref{chem-aff}).}
\\

We notice that the function (\ref{G_tilde}) takes the unit value if $\xi=\eta_{\alpha}=0$ so that
the cumulant generating function vanishes with the counting parameters:
\be
{\cal Q}(0,0;{\cal B}) = {\cal Q}(A_E,A_{\alpha};-{\cal B})=0 \; .
\label{Q=0}
\ee

The current fluctuation theorem (\ref{CFT}) can be extended to the case of systems with more than two reservoirs \cite{AGMT09}. If all the $r$ reservoirs are coupled together, the total energy as well as the total numbers of particles are conserved in the transport process so that there are $(c+1)$ constants of motion.  The grand-canonical equilibrium state of each reservoir is specified by one temperature and $c$ chemical potentials.  One of the $r$ reservoirs can be taken as a reference with respect to which the affinities are defined.  Consequently, the nonequilibrium steady states are specified by $p=(c+1)(r-1)$ different affinities and so many independent currents may flow across the open subsystem in between the reservoirs.

Some systems may be composed of several separated circuits between which no current is flowing.  This is the case for instance in quantum dots monitored by a secondary circuit with a quantum point contact \cite{GLSSISEDG06,GLSSIEDG09,FHTH06}.  Both circuits are only coupled by the Coulomb interaction so that there is no electron transfer between them.  In such systems, there are more independent affinities and currents.  If a system composed of $r$ reservoirs is partitioned into $d$ disconnected circuits each containing at least two reservoirs, a reference reservoir should be taken in each separate circuit.  Therefore, $(r-d)$ reservoirs are controlling the $(c+1)$ possible currents.  The nonequilibrium steady states are thus specified by $p=(c+1)(r-d)$ different affinities and so many independent currents can flow in such systems.

\section{The current fluctuation theorem and response theory}
\label{Resp}

If we collect together all the independent affinities and counting parameters as
\bea
&& {\bf A} = \{ A_{jE}, A_{j\alpha} \} \\
&& \pmb{\lambda} = \{ \xi_j-\xi_k, \eta_{j\alpha}- \eta_{k\alpha} \}
\eea
where $k$ stands for the indices of the $d$ reference reservoirs, the current fluctuation theorem can be expressed as
\be
{\cal Q}(\pmb{\lambda},{\bf A};{\cal B}) = {\cal Q}({\bf A}-\pmb{\lambda},{\bf A};-{\cal B})
\label{CFT-bis}
\ee
where we have introduced explicitly the dependence of the generating function on the affinities ${\bf A}$ which specify the nonequilibrium steady state of the open system.
The maximum number of independent affinities and counting parameters is equal to $p=(c+1)(r-d)$.  If the total system is isothermal, there is no energy current and the number of independent affinities and counting parameters is equal to $p=c \times (r-d)$ where $c$ is the number of different particle species. 

All the statistical cumulants of the energy and particles transferred between the reservoirs are obtained by taking successive derivatives with respect to the counting parameters and setting afterwards all these parameters equal to zero.  The mean values of the currents are given by the first derivatives, the diffusivities or second cumulants by the second derivatives, and similarly for the third or higher cumulants:
\bea
J_{\alpha}({\bf A};{\cal B}) &\equiv& \frac{\partial {\cal Q}}{\partial
\lambda_{\alpha}}({\bf 0},{\bf A};{\cal B}) \label{av_J} \\
D_{\alpha\beta}({\bf A};{\cal B}) &\equiv& - \frac{1}{2} \frac{\partial^2 {\cal Q}}{\partial
\lambda_{\alpha}\partial \lambda_{\beta}}({\bf 0},{\bf A};{\cal B}) \label{D}\\
C_{\alpha\beta\gamma}({\bf A};{\cal B}) &\equiv& \frac{\partial^3 {\cal Q}}{\partial
\lambda_{\alpha}\partial \lambda_{\beta}\partial \lambda_{\gamma}}({\bf 0},{\bf A};{\cal B})
\label{C} \\
&\vdots& \nonumber
\eea
with $\alpha,\beta,...=1,2,...,p$.  All these cumulants characterize the full counting statistics of the coupled fluctuating currents in the nonequilibrium steady state associated with the affinities $\bf A$ in the external magnetic field $\cal B$.

At equilibrium, the mean currents vanish with the affinities.  Therefore, we may expect that the mean currents can be expanded in powers of the affinities close to equilibrium:
\be
J_{\alpha} =
\sum_{\beta} L_{\alpha,\beta}\,  A_{\beta} + \frac{1}{2}
\sum_{\beta,\gamma} M_{\alpha,\beta\gamma}\,  A_{\beta}
A_{\gamma}
+ \cdots 
\label{dfn-currents}
\ee
which defines the {\it response coefficients}:
\bea
L_{\alpha,\beta}({\cal B}) &\equiv& 
\frac{\partial^2 {\cal Q}}{\partial \lambda_{\alpha}\partial A_{\beta}}({\bf 0},{\bf 0};{\cal B}) \; ,\\
M_{\alpha,\beta\gamma}({\cal B}) &\equiv& 
\frac{\partial^3 {\cal Q}}{\partial \lambda_{\alpha}\partial A_{\beta}\partial A_{\gamma}}({\bf 0},{\bf 0};{\cal B}) \; ,\label{M}\\
&& \nonumber \\
&\vdots& \nonumber
\eea

We observe that the cumulants are given by successive derivatives with respect to the counting parameters.  On the other hand, the response coefficients are given by one derivative with respect to a counting parameter and further successive derivatives with respect to the affinities.  The remarkable result is that differentiating the symmetry relation (\ref{CFT-bis}) with respect to the affinities leads to derivatives with respect to the counting parameters.  In this way, the current fluctuation theorem implies fundamental relationships between the cumulants and the response coefficients.

Considering all the second derivatives of the generating function with respect to the counting parameters and the affinities, it is possible to deduce the Casimir-Onsager reciprocity relations
\be
L_{\alpha,\beta}({\cal B}) =  L_{\beta,\alpha}(-{\cal B}) \; ,
\label{Casimir}
\ee
as well as the identities
\be
L_{\alpha,\beta}({\cal B}) + L_{\beta,\alpha}({\cal B}) = 2 \, D_{\alpha\beta}({\bf 0};{\cal B})
\label{FD-thm}
\ee
which relate the linear response coefficients to the diffusivities at equilibrium where ${\bf A}={\bf 0}$.

In the absence of magnetic field ${\cal B}=0$, we recover the Onsager reciprocity relations
\be
L_{\alpha,\beta}(0) =  L_{\beta,\alpha}(0) \; ,
\ee
as well as the expression of the fluctuation-dissipation theorem according to which the linear response coefficients are equal to the diffusivities of the currents around equilibrium:
\be
L_{\alpha,\beta}(0) = D_{\alpha\beta}({\bf 0};0) \; .
\label{L-D}
\ee

Going to the third derivatives of the symmetry relation (\ref{CFT-bis}), the third cumulants (\ref{C}) characterize a magnetic asymmetry at equilibrium where ${\bf A}={\bf 0}$ and they are related to the responses of the diffusivities with respect to the affinities according to
\be
C_{\alpha\beta\gamma}({\bf 0};{\cal B}) = - C_{\alpha\beta\gamma}({\bf 0};-{\cal B}) = 2\, 
\frac{\partial D_{\alpha\beta}}{\partial A_{\gamma}}({\bf 0};{\cal B}) -2 \, 
\frac{\partial D_{\alpha\beta}}{\partial A_{\gamma}}({\bf 0};-{\cal B}) \; .
\label{C0-D1}
\ee
Moreover, the following relations hold for the nonlinear response coefficients at second order in the perturbations with respect to equilibrium \cite{AGMT09}
\bea
&& M_{\alpha,\beta\gamma}({\cal B})+M_{\alpha,\beta\gamma}(-{\cal B}) = 2\, \frac{\partial D_{\alpha\beta}}{\partial A_{\gamma}}({\bf 0};{\cal B}) +2\,
\frac{\partial D_{\alpha\gamma}}{\partial A_{\beta}}({\bf 0};-{\cal B}) \; , \label{M-D1}\\
&& M_{\alpha,\beta\gamma}({\cal B})+M_{\beta,\gamma\alpha}({\cal B})+M_{\gamma,\alpha\beta}({\cal B}) = 2 \left(\frac{\partial D_{\beta\gamma}}{\partial A_{\alpha}}+
\frac{\partial D_{\gamma\alpha}}{\partial A_{\beta}} +
\frac{\partial D_{\alpha\beta}}{\partial A_{\gamma}} - 
\frac{1}{2} \, C_{\alpha\beta\gamma}\right)_{{\bf A}={\bf 0};\,{\cal B}} \; .
\label{M-D1-C0}
\eea

In the absence of magnetic field ${\cal B}=0$, the magnetic asymmetry disappears because the third cumulants vanish, $C_{\alpha\beta\gamma}({\bf 0};0)=0$, and the response coefficients are given in terms of the diffusivities as follows \cite{AG04,AG06JSM,AG07JSM}
\be
M_{\alpha,\beta\gamma}(0) = \left(\frac{\partial D_{\alpha\beta}}{\partial A_{\gamma}}+
\frac{\partial D_{\alpha\gamma}}{\partial A_{\beta}}\right)_{{\bf A}={\bf 0};\,{\cal B}=0} \; .
\ee
These relations are the generalizations of Eq.~(\ref{L-D}) to nonlinear response coefficients.

Such relations, which can be extended to higher orders as well \cite{AGMT09,AG04,AG06JSM,AG07JSM},
find their origin in the microreversibility expressed by the current fluctuation theorem (\ref{CFT-bis}).

\section{The case of independent particles}
\label{Indep}

In order to illustrate the previous results, let us consider systems with independent fermionic particles ruled by the Hamiltonian:
\be
\hat H = \sum_{\sigma=\pm} \int  d{\bf r} \, \hat\psi_{\sigma}^{\dagger}({\bf r}) \, \hat h \, \hat\psi_{\sigma}({\bf r})
\label{H_space}
\ee
written in terms of the the anticommuting field operators $\hat\psi_{\sigma}({\bf r})$
and the one-particle Hamiltonian operator:
\be
\hat h = - \frac{\hbar^2}{2m} \, \nabla^2 + u({\bf r}) \qquad\mbox{with}\qquad {\bf r}=(x,y,z)\; .
\ee

The confining potential $u({\bf r})$ is minimum in the conducting region of the circuit and presents a barrier at the borders of this region.  Such confining potentials can be fabricated by nanolithography on semiconducting heterojunctions and various shapes can be given to the circuit \cite{FGB09}.  
If the temperature is of the order of a few Kelvins or less, only the electrons around the Fermi energy are transported.  The de Broglie wavelength can be larger than the interwall distance so that the circuit behaves as an electronic waveguide.

\begin{figure}[h]
\begin{center}
\includegraphics[scale=0.3]{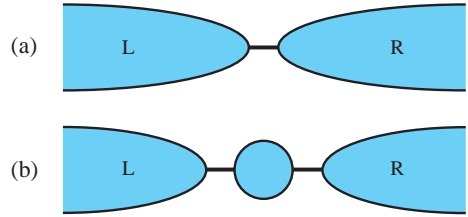}
\caption{Schematic representation of different electronic circuits: (a) quantum point contact between two reservoirs; (b)~quantum dot in tunneling contact with two reservoirs.}
\label{fig3}
\end{center}
\end{figure}

A quantum point contact (QPC) is a circuit with a barrier separating two semi-infinite waveguides, which form two reservoirs (see Fig.~\ref{fig3}a). The barrier is a bottleneck for the transport of electrons from one reservoir to the other.  This bottleneck presents a saddle point in the potential energy landscape. If the Fermi energy is lower than the energy of the saddle point, the transport proceeds by quantum tunneling through the barrier.  If the Fermi energy is larger, the electronic waves undergo direct scattering on the obstacle formed by the barrier.  Different models can be envisaged for the potential of the QPC.  At large distance from the barrier, electrons can be assumed to propagate in a waveguide with a potential $u({\bf r})=u_{\bot}(y,z)$, which is invariant under translations in the $x$-direction and confining in the transverse $y$- and $z$-directions.  The propagation modes in the infinite waveguide are given by the eigenstates of the one-electron Hamiltonian:
\be
\hat h_0 \, \psi_{k_xn_yn_z}({\bf r})  = \varepsilon_{k_xn_yn_z} \, \psi_{k_xn_yn_z}({\bf r})
\label{mode}
\ee
where $k_x$ is the wavenumber in the $x$-direction of propagation while $n_y$ and $n_z$ are the quantum numbers of the modes in the transverse directions.  Each of these modes is a possible channel of conduction, which opens at the energy threshold given by the minimum of the energy band $\varepsilon_{k_xn_yn_z}$ at $k_x=0$.

The propagation modes $\psi_{k_xn_yn_z}({\bf r})$ are scattered by the barrier.
If the potential can be assumed to be the sum of the transverse and barrier potentials, $u({\bf r})=u_{\Vert}(x)+u_{\bot}(y,z)$, the problem is unidimensional for the passage of the barrier.  If the incoming wavefunction is a plane wave in the positive $x$-direction, the scattering generates reflected and transmitted waves:
\be
\psi_k(x) = \left\{
\begin{array}{ll}
\frac{1}{\sqrt{2\pi}} \, \left( {\rm e}^{ikx} + r_k \,  {\rm e}^{-ikx}\right) &\qquad\mbox{for} \quad x<0 \; , \\
\frac{1}{\sqrt{2\pi}} \, t_k \,  {\rm e}^{ikx} &\qquad\mbox{for} \quad x>0 \; , 
\end{array}
\right.
\ee
where $r_k$ and $t_k$ denote the reflection and transmission amplitudes while $k=k_x$ is the wavenumber.  At the energy $\varepsilon=\varepsilon(k)$, the outgoing waves are related to the incoming waves by the unitary scattering matrix according to
\be
\Psi_{\rm out} = \hat S(\varepsilon)\,  \Psi_{\rm in}
\ee
where $\Psi_{\rm in}$ and $\Psi_{\rm out}$ denote vectors with two components.  The first component is the wave amplitude in the left-hand reservoir and the second in the right-hand reservoir.
Therefore, the unitary scattering matrix of the one-dimensional barrier is given by
\be
\hat S(\varepsilon) = \left(
\begin{array}{cc}
r_k & t_k \\
t_k & r_k 
\end{array}
\right) \qquad\mbox{such that}\qquad \hat S^{\dagger}(\varepsilon)\, \hat S(\varepsilon) = 1 \; .
\label{S-matrix}
\ee
The unitarity of the scattering matrix expresses the conservation of probability and, in particular, the fact that the transmission probability $T(\varepsilon)=\vert t_k\vert^2$ and the reflection probability $R(\varepsilon)=\vert r_k\vert^2$ add to the unit value: $T(\varepsilon)+R(\varepsilon)=1$.  Here, the scattering matrix is $2\times 2$ for every transverse mode because the potential defines a problem that is separable into the longitudinal wave and the transverse modes.  This would not be the case if the potential was no longer the sum of two potentials in perpendicular directions.  In more complicated circuits such as billiards, it is known that the scattering matrix is infinite and couples together the different transverse modes \cite{GAON94,NH04}.

In the case of separable potentials, every mode is scattered independently of the other ones and the scattering is thus characterized by the transmission probability $T(\varepsilon)=\vert t_k\vert^2$, which depends on the wavenumber $k=k_x$ or, equivalently, on the corresponding energy $\varepsilon=\varepsilon(k)$ of the electron in the longitudinal direction.  A well-known example is the inverted parabolic barrier $u(x)=u_0-m\lambda^2x^2/2$, for which the transmission probability is given by \cite{NB09,B90}
\be
T(\varepsilon)= \left[1+\exp\left(-2\pi \,\frac{\varepsilon-u_0}{\hbar\lambda}\right)\right]^{-1} \; .
\label{T-parabolic}
\ee
The transmission probability converges to the unit value at high energy well above the height of the barrier.  For energies lower than the height of the barrier, the transmission proceeds by tunneling.  The broader the barrier, the smaller the transmission probability.  Since the inverted parabolic potential is unbounded from below, the transmission probability (\ref{T-parabolic}) remains positive for $\varepsilon\leq 0$ and vanishes only in the limit $\varepsilon\to-\infty$.  For potentials which are vanishing at large distances, the transmission probability is strictly equal to zero if $\varepsilon\leq 0$.

A further example of electronic circuit is composed of a quantum dot between two reservoirs, as depicted in Fig.~\ref{fig3}b.  A quantum dot is an artificial atom in which several electrons are confined on quasi discrete energy levels \cite{FGB09,NB09}.  In general, there exist several electronic energy levels but the transmission can be observed around a single electronic level.  If the tunneling to the reservoirs is small enough, the energy levels are observed as resonances in the transmission probability.  For quantum circuits such as classically chaotic billiards, the scattering resonances may form an irregular spectrum in the complex plane of the wavenumber or the energy and semiclassical methods become useful \cite{GAON94,NH04}.

An important remark is that the Hamiltonian operator (\ref{H_space}) can also be written in the form:
\be
\hat H= \hat H_{\rm L} + \hat H_{\rm R} + \hat V_0
\ee
where $\hat H_{\rm L}$ and $\hat H_{\rm R}$ are the Hamiltonian operators of the left- and right-hand reservoirs, while $\hat V_0$ is the interaction between the reservoirs, this interaction being at the origin of the scattering.

For these systems, we consider the function (\ref{G_tilde}) in the absence of magnetic field and for one species of particles, namely, the electrons.  This function has the following form:
\be
\tilde {\cal G}_t(\xi,\eta) = {\rm tr}\, \hat\rho_0 \, {\rm e}^{-\hat A_t} \, {\rm e}^{\hat A_0}
\label{G_tilde_bis}
\ee
with
\be
\hat A = \frac{\xi}{2}(\hat H_{\rm R}-\hat H_{\rm L})+\frac{\eta}{2}(\hat N_{\rm R}-\hat N_{\rm L})
\ee
and
\be
\hat A_t = {\rm e}^{i\hat H t} \, \hat A \, {\rm e}^{-i\hat H t}
\ee
where $\hat H_1=\hat H_{\rm R}$ and $\hat H_2=\hat H_{\rm L}$, in order to define the affinities (\ref{th-aff}) and (\ref{chem-aff}) corresponding to currents from the left-hand reservoir to the right-hand one.

According to Eq.~(\ref{rho_0}), the initial density operator is given by
\be
\hat\rho_0=\frac{{\rm e}^{-\hat B}}{{\rm tr}\, {\rm e}^{-\hat B}}
\qquad\qquad\mbox{with}\qquad\qquad
\hat B = \beta_{\rm L} (\hat H_{\rm L} -\mu_{\rm L} \hat N_{\rm L}) + \beta_{\rm R} (\hat H_{\rm R} -\mu_{\rm R} \hat N_{\rm R})
\ee
in terms of the inverse temperatures, the chemical potentials, and the particle-number operators, $\hat N_{\rm L}$ and $\hat N_{\rm R}$, for both reservoirs.  Since $[\hat A,\hat B]=0$, the function (\ref{G_tilde_bis}) can be written as
\be
\tilde {\cal G}_t(\xi,\eta) = \frac{{\rm tr}\, {\rm e}^{-\hat A_t} \, {\rm e}^{\hat A-\hat B}}{{\rm tr}\, {\rm e}^{-\hat B}} \; .
\label{G_tilde_ter}
\ee

In systems with independent particles, many-particle operators such as the Hamiltonian and particle-number operators are of the form
\be
\hat X = \sum_{k \, l} x_{kl} \, \hat c_{k}^{\dagger} \, \hat c_{l} = \Gamma(\hat x)
\label{map}
\ee
where $\hat x = (x_{kl})$ denotes the corresponding one-particle operator, while
$\hat c_{k}^{\dagger}$ and $\hat c_k$ are the anticommuting creation-annihilation operators in the one-particle state $k$.  The correspondence between one-particle and many-particle operators is thus established by the mapping $\Gamma$ defined by Eq.~(\ref{map}).
In this framework, Klich has shown that the trace of products of exponential functions of many-particle operators can be expressed as appropriate determinants involving the one-particle operators \cite{K03,ABGL08}.  For fermions, Klich's formula reads
\be
{\rm tr} \, {\rm e}^{\hat X} \, {\rm e}^{\hat Y} =  {\rm tr} \, {\rm e}^{\Gamma(\hat x)} \, {\rm e}^{\Gamma(\hat y)} =  \det \left( 1 + {\rm e}^{\hat x} \, {\rm e}^{\hat y} \right) \; .
\ee
Applying Klich's formula to the function (\ref{G_tilde_ter}), this latter becomes
\be
\tilde {\cal G}_t(\xi,\eta) = \det\left[\left(1+ {\rm e}^{-\hat b}\right)^{-1}\left(1 +{\rm e}^{-\hat a_t} \, {\rm e}^{\hat a-\hat b}\right)\right] 
= \det\left[1+ \hat f \left({\rm e}^{-\hat a_t} \, {\rm e}^{\hat a} - 1 \right)\right] 
\label{G_tilde_4}
\ee
with the operator: 
\be
\hat f = \frac{1}{{\rm e}^{\hat b} + 1} = \frac{1}{{\rm e}^{\beta_{\rm L} (\hat h_{\rm L} -\mu_{\rm L} \hat n_{\rm L}) + \beta_{\rm R} (\hat h_{\rm R} -\mu_{\rm R} \hat n_{\rm R})
} + 1} 
\ee
of the Fermi-Dirac distributions in the reservoirs.

In the long-time limit, the unitary scattering operator is defined as
\be
\hat S = \lim_{t\to\infty} {\rm e}^{i\hat h_0t/2} \,  {\rm e}^{-i\hat ht} \, {\rm e}^{i\hat h_0t/2} 
\ee
in terms of the full Hamiltonian $\hat h$ and the non-interacting Hamiltonian $\hat h_0=\hat h_{\rm L}+\hat h_{\rm R}$ which commutes with the operators $\hat a$ and $\hat b$.
Since
\be
\hat a_t = {\rm e}^{i\hat ht} \, \hat a \, {\rm e}^{-i\hat ht} 
\ee
we have that
\be
{\rm e}^{-\hat a_t} \simeq {\rm e}^{i\hat h_0t/2} \, \hat S^{\dagger} \, {\rm e}^{-\hat a}\, \hat S \, {\rm e}^{-i\hat h_0t/2}
\ee
in the long-time limit.  Using the commutativity of $\hat h_0$ with $\hat a$ and $\hat f$, the function (\ref{G_tilde_4}) can be written as
\be
\tilde {\cal G}_t(\xi,\eta) = \det\left[1+ \hat f \left(\hat S^{\dagger} \, {\rm e}^{-\hat a}\, \hat S \, {\rm e}^{\hat a} - 1 \right)\right] 
\label{G_tilde_5}
\ee

For this formula to make sense, we have to notice that the determinant should be taken over an appropriate discrete set of electronic states forming a quasi continuum in the long-time limit.  Indeed, the quantity in the left-hand side of Eq.~(\ref{G_tilde_5}) concerns the whole many-particle system although the expression in the right-hand side concerns single independent electrons flowing one by one across the system.  Every one-particle operator can be decomposed on the eigenstates of the non-interacting Hamiltonian $\hat h_0=\hat h_{\rm L}+\hat h_{\rm R}$.  This is the case in particular for the scattering operator:
\be
\hat S = \int d\varepsilon \, \hat S(\varepsilon) \, \delta(\varepsilon - \hat h_0)
\ee
where $\hat S(\varepsilon)$ is the $2\times 2$ scattering matrix (\ref{S-matrix}) acting on the two wave amplitudes with opposite wavenumbers $\pm k$ at the given energy $\varepsilon=\varepsilon(k)$.  Accordingly, the function (\ref{G_tilde_5}) becomes a product over all the relevant single-electron states $\{\varepsilon,\sigma\}$ of the determinants of the corresponding operators at this particular energy:
\be
\tilde {\cal G}_t(\xi,\eta) = \prod_{\varepsilon,\sigma} \det\left\{1+ \hat f(\varepsilon) \left[\hat S^{\dagger}(\varepsilon) \, {\rm e}^{-\hat a(\varepsilon)}\, \hat S(\varepsilon) \, {\rm e}^{\hat a(\varepsilon)} - 1 \right]\right\}
\label{G_tilde_6}
\ee
where the matrix containing the Fermi-Dirac distributions of the left- and right-hand reservoirs is given by
\be
\hat f(\varepsilon) = 
\left(
\begin{array}{cc}
f_{\rm L}(\varepsilon) & 0 \\
0 & f_{\rm R}(\varepsilon) 
\end{array}
\right)
\qquad\mbox{with}\qquad
f_{\rm L,R}(\varepsilon) = \frac{1}{{\rm e}^{\beta_{\rm L,R}(\varepsilon-\mu_{\rm L,R})} + 1} 
\ee
and 
\be
{\rm e}^{\hat a(\varepsilon)} = 
\left(
\begin{array}{cc}
{\rm e}^{-(\varepsilon\, \xi + \eta)/2} & 0 \\
0 & {\rm e}^{+(\varepsilon\, \xi + \eta)/2}
\end{array}
\right) \; .
\ee

In order to define the cumulant generating function with Eq.~(\ref{Q-dfn}), the logarithm is taken of the function (\ref{G_tilde_6}).  The logarithm of the product over the relevant states $\{\varepsilon,\sigma\}$ gives a sum of logarithms.  For a process lasting over the lapse of time $t$, the spacing between the relevant energies $\{\varepsilon\}$ is equal to $\Delta\varepsilon=2\pi\hbar /t$. Since these relevant energies form a quasi continuum, the sum is replaced by an integral in the long-time limit.  Moreover, the spin orientation takes the two values $\sigma=\pm$ for electrons. For spin one-half particles such as electrons, we thus find that
\be
\lim_{t\to\infty} \, \frac{1}{t} \, \sum_{\varepsilon,\sigma} \, (\cdot) = 2_s \int \frac{d\varepsilon}{2\pi\hbar} \, (\cdot)
\ee
where $2_s$ denotes the factor two due to the electron spin \cite{NB09}.

After evaluating the determinant, the cumulant generating function (\ref{Q-dfn}) is finally given by
\be
{\cal Q}(\xi,\eta) = - 2_s \int \frac{d\varepsilon}{2\pi\hbar} \, \ln\left\{ 1 + T(\varepsilon) \left[ f_{\rm L} (1-f_{\rm R}) \left({\rm e}^{-\varepsilon\xi-\eta}-1\right) + f_{\rm R} (1-f_{\rm L}) \left({\rm e}^{\varepsilon\xi+\eta}-1\right)\right]\right\}
\label{Q-LL}
\ee
for independent electrons.  Therefore, we have recovered the Levitov-Lesovik formula for the full counting statistics of electron quantum transport \cite{LL93,LLL96}.  The symmetry of the fluctuation theorem is satisfied:
\be
{\cal Q}(\xi,\eta) = {\cal Q}(A_E-\xi,A_N-\eta) 
\label{FT-LL}
\ee
with respect to the thermal and chemical affinities 
\bea
&& A_E =  \beta_{\rm R}-\beta_{\rm L} \; , \label{A_E}\\
&& A_N =  \beta_{\rm L}\, \mu_{\rm L} -\beta_{\rm R} \, \mu_{\rm R} \; , \label{A_N}
\eea
defined by Eqs.~(\ref{th-aff}) and (\ref{chem-aff}) if the reservoir No.~1 (respectively No.~2) is identified with the right-hand (respectively left-hand) reservoir.

\begin{figure}[h]
\begin{center}
\includegraphics[scale=0.4]{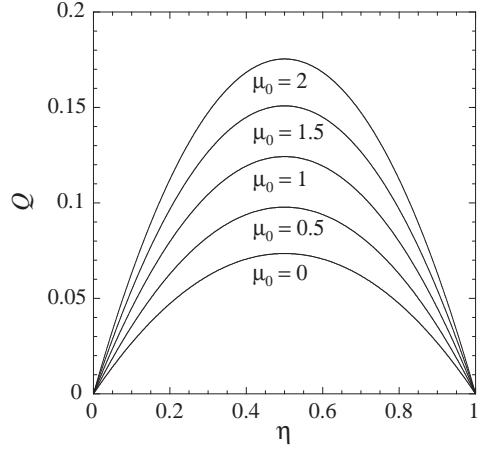}
\caption{The cumulant generating function (\ref{Q-LL}) of the particle current versus the counting parameter $\eta$ for the inverted parabolic barrier $u(x)=u_0-m\lambda^2x^2/2$ with the transmission probability (\ref{T-parabolic}). The temperature difference and, thus, the thermal affinity (\ref{A_E}) are vanishing.  In this regard, the generating function is considered at $\xi=0$.  The chemical potentials of the left- and right-hand reservoirs are given by $\mu_{\rm L,R}=\mu_0\pm eV/2$ with $eV=1$ and $\mu_0=0,0.5,1,1.5,2$.  The other parameter values are $u_0=1$, $\hbar\lambda=3$, and $\beta=(k_{\rm B}T)^{-1}=1$.  Planck's constant is taken as $h=2_s$.  The plot confirms the symmetry $\eta\to A_N-\eta$ of the fluctuation theorem with respect to the chemical affinity $A_N=\beta(\mu_{\rm L}-\mu_{\rm R})=\beta eV = 1$.}
\label{fig4}
\end{center}
\end{figure}

The cumulant generating function (\ref{Q-LL}) with $\xi=A_E=0$ is depicted in Fig.~\ref{fig4} for the inverted parabolic barrier with the transmission probability (\ref{T-parabolic}). We see that the generating function has indeed the symmetry $\eta\to A_N-\eta$ of Eq.~(\ref{FT-LL}).  The generating function is depicted for several values of the mean chemical potential $\mu_0=(\mu_{\rm L}+\mu_{\rm R})/2$ of both reservoirs. We observe that the generating function increases with the mean chemical potential $\mu_0$ because the transmission probability (\ref{T-parabolic}) does so.

The generating function given by the Levitov-Lesovik formula can be interpreted in terms of an exclusion process ruled by a trinomial distribution \cite{RDD05}.  Suppose that, at every relevant energy, three types of random events may happen for an electron:  it may move from the left- to the right-hand reservoir with the probability $p_+$, in the other direction with the probability $p_-$, and stay in some reservoir with the probability $p_0=1-p_+ -p_-$.  The probability that $N_0$ electrons do not move, $N_+$ electrons move to the right-hand reservoir, and $N_-$ to the left-hand reservoir during the time interval $t$ is given by the trinomial distribution:
\be
\tilde P_{N_0N_+N_-}= \frac{N!}{N_0!\, N_+!\, N_-!} \ p_0^{N_0} \, p_+^{N_+} \, p_-^{N_-} \qquad \mbox{with}\qquad N=N_0+N_++N_- \; .
\ee
For this random process, the generating function is defined as
\be
\tilde {\cal Q}(\eta) = \lim_{t\to\infty} \, - \frac{1}{t}\, \ln \langle {\rm e}^{-\eta \, \Delta N} \rangle_t
\ee
where $\Delta N = N_+ - N_-$ is the total number of electrons that have been transported from the left- to the right-hand reservoir during the time interval $t$.  Taking the average over the trinomial probability distribution, we find that
\be
\tilde {\cal Q}(\eta)= - r \, \ln \left[ 1 + p_+\, ({\rm e}^{-\eta}-1)+ p_-\, ({\rm e}^{\eta}-1)\right] 
\label{Q-trinomial}
\ee
where $r=\lim_{t\to\infty}(N/t)$ is the attempt frequency, i.e., the mean number of attempted electron transfers per unit time \cite{NB09}.  We observe that the special form of the generating function (\ref{Q-LL}) is indeed obtained at every relevant energy if we take $p_+=T(\varepsilon)f_{\rm L}(1-f_{\rm R})$ and $p_-=T(\varepsilon)f_{\rm R}(1-f_{\rm L})$, as it should.  The generating function of the trinomial process has the symmetry
\be
\tilde {\cal Q}(\eta) = \tilde {\cal Q}(A_N-\eta) \qquad \mbox{with} \qquad A_N = \ln \frac{p_+}{p_-} \; .
\ee
This provides a stochastic interpretation for the quantum transport process described by the Levitov-Lesovik formula~(\ref{Q-LL}).

Many known results for quantum transport can be deduced in this framework.
In particular, the mean currents are given by
\bea
&&  J_N = \frac{\partial{\cal Q}}{\partial\eta}(0,0) = 2_s \int \frac{d\varepsilon}{2\pi\hbar} \, T(\varepsilon) \left( f_{\rm L}-f_{\rm R}\right) \; , \\
&& J_E  = \frac{\partial{\cal Q}}{\partial\xi}(0,0)= 2_s \int \frac{d\varepsilon}{2\pi\hbar} \, \varepsilon \, T(\varepsilon) \left( f_{\rm L}-f_{\rm R}\right) \; ,
\eea
which are the well-known Landauer formula for the particle current and the equivalent formula for the energy current \cite{SI86}.
On the other hand, the noise is characterized by the diffusivities:
\be
D_{NN} = -\frac{1}{2}\, \frac{\partial^2{\cal Q}}{\partial\eta^2}(0,0) = D_0\; , \qquad
D_{EN} = -\frac{1}{2}\, \frac{\partial^2{\cal Q}}{\partial\eta\partial\xi}(0,0) = D_1 \; ,\qquad
D_{EE} = -\frac{1}{2}\, \frac{\partial^2{\cal Q}}{\partial\xi^2}(0,0) = D_2 \; ,
\ee
with
\be
D_m = \int \frac{d\varepsilon}{2\pi\hbar} \, \varepsilon^m  \, T(\varepsilon) \left[ f_{\rm L}(1-f_{\rm R})+f_{\rm R}(1-f_{\rm L})-T(\varepsilon) (f_{\rm L}-f_{\rm R})^2 \right] \qquad\mbox{for}\qquad m=0,1,2
\ee
as reported in Refs.~\cite{NB09,TT05,BB00}.
The power spectrum of the electric noise is defined as
\be
S(\omega) = \int dt \, {\rm e}^{i\omega t} \langle \Delta \hat I(t)\, \Delta \hat I(0) + \Delta \hat I(0)\, \Delta \hat I(t)\rangle
\label{noise_power}
\ee
with $\Delta \hat I(t)= \hat I(t)-\langle\hat I(t)\rangle$ \cite{BB00}.  Since $\hat I = e \hat J_N$, the zero-frequency limit of the noise power is thus related to the corresponding diffusivity by 
\be
S(\omega=0)=4e^2 D_{NN}\; .
\label{S-D}
\ee

Close to equilibrium, the mean currents can be expanded in powers of the affinities (\ref{A_E}) and (\ref{A_N}). However, alternative expansions can be considered in terms of the potential difference $V$ and the temperature difference $\Delta T$ between the left- and right-hand reservoirs at the temperatures $T_{\rm L,R} = T_0 \pm \Delta T/2$ and chemical potentials $\mu_{\rm L,R} = \mu_0 \pm eV/2$.
If we introduce the electric and heat currents as
\bea
&&  I  = e \,  J_N  \; , \\
&&  J_Q  =  J_E  - \mu_0 \,  J_N  \; ,
\eea
with the electric charge $e$, the alternative expansions around equilibrium read:
\bea
&&  I  \simeq G \, V + L_1 \, \frac{\Delta T}{T_0}  \; , \\
&& J_Q  \simeq L_1 \, V + L_2 \, \frac{\Delta T}{T_0}  \; ,
\eea
if higher-order terms are neglected.  In this linear approximation,
$G=L_0$ is the electric conductance, $L_1$ the thermo-electric coupling coefficient,
and $L_2$ the thermal conductance, which are given by \cite{SI86}
\be
L_m = 2_s \, e^{2-m} \int \frac{d\varepsilon}{2\pi\hbar}  \, (\varepsilon-\mu_0)^m \; T(\varepsilon) \left( - \frac{\partial f}{\partial\varepsilon}\right)_0 \qquad\mbox{with}\qquad m=0,1,2 \; .
\label{linear-coeff}
\ee
We notice that the Onsager reciprocity relation is satisfied.

As the example of Eq.~(\ref{T-parabolic}) shows, the transmission probability tends to the unit value for energies higher than the barrier in every open channel corresponding to a propagation mode. In this limit where $T(\varepsilon)=1$, the electric conductance  given by Eq.~(\ref{linear-coeff}) with $m=0$ takes the universal value:
\be
G_{\rm q}=\frac{2_s \, e^2}{2\pi\hbar} = \frac{2 \, e^2}{h} \; .
\ee
If several channels are open at the Fermi energy $\varepsilon_{\rm F}$,  the zero-temperature conductance is given by
\be
G = \frac{2 \, e^2}{h} \sum_n  T_n(\varepsilon_{\rm F}) 
\ee
where $T_n(\varepsilon)$ is the transmission probability of the channel corresponding to the mode $n=(n_y,n_z)$ in Eq.~(\ref{mode}). As the Fermi energy increases, the channels open successively and their transmission probability tends to the unit value, which explains the phenomenon of conductance quantization \cite{I97,FGB09,NB09,BB00,vWvHBWKvdMF88}.

In summary, the results obtained in the previous sections allow us to deduce systematically many well-known results on quantum transport such as the Levitov-Lesovik and the Landauer formulae in the scattering approach, which is appropriate to treat coherent quantum transport.

Similar considerations also apply to boson transport \cite{SD07,HEM07} as well as to many other quantum transport processes in physics or chemistry where the theory of full counting statistics extends and complements standard scattering theory and reaction rate theory.

We notice that these quantum transport properties can also be deduced in the so-called Keldysh formalism, which has several conceptual advantages \cite{R07,NB09,FGB09}.  On the one hand, it formulates transport in terms of quantum fields, allowing in principle to treat the spatial dependence of local properties such as densities or current densities. On the other hand, the Keldysh formalism provides the systematic introduction of all the moments of the quantum fields, starting from the nonequilibrium Green's functions, which is suitable for perturbative calculations.
Several nonequilibrium Green's functions are introduced corresponding to different time contours, which come with the doubling of the space of quantum states from pure states to statistical mixtures described by density operators.
An equivalent theory is the Liouville-space formalism developed by Mukamel and coworkers \cite{M95,HM08,EHM09}, which starts from the quantum Liouville superoperator, i.e., the generator of von Neumann's equation (\ref{vN_eq}).  Another equivalent theory is provided by the formalism of thermofield dynamics, which also takes into account the need to double of the state space for the description of nonequilibrium systems \cite{U95}.  The latter formalism has recently been applied to study electron transport in molecular junctions \cite{K09,DK11}.

\section{Time-reversal symmetry relations in the master-equation approach}
\label{Master}

\subsection{Current fluctuation theorem for stochastic processes}

The master-equation approach allows us to establish connections with the theory of stochastic processes. The first step of this approach consists in identifying the relevant coarse-grained states of the system or subsystem of interest.  The second step usually proceeds with a perturbative calculation with respect to some small coupling parameter $\lambda$, which typically controls the time scale separation between the slow time evolution of the coarse-grained states and the fast dynamics of the other degrees of freedom.  The Hamiltonian may have the form
\be
\hat H = \hat H_0 + \lambda \hat V
\ee
where the unperturbed Hamiltonian $\hat H_0$ leaves invariant the coarse-grained states and the perturbation $\lambda\hat V$ causes the interaction between the coarse-grained states and the other degrees of freedom. Without a clear separation of time scales, the time evolution of the coarse-grained states continues to keep the memory of the past and is non-Markovian.  In quantum-mechanical systems, non-Markovian behavior can manifest itself as a slippage of initial conditions on the time scale of the fast degrees of freedom in the early time evolution of a subsystem in contact with a reservoir.  The slippage of initial conditions allows the density operator to remain positive definite \cite{GN99a,GN99b}.  Thereafter, the density operator of the subsystem follows a time evolution which is essentially Markovian on the long time scale characterizing the interaction of the coarse-grained states with the fast degrees of freedom. We notice that, although the subsystem density operator evolves slowly, the individual realizations present quantum jumps between the coarse-grained states.  These jumps occur over the short time scale of the fast degrees of freedom and the dwell times between the jumps are of the order of the long time scale.  Indeed, these dwell times are inversely proportional to the transition rates, which are of the order of $\lambda^2$ according to second-order perturbation theory.

In transport problems where the subsystem of interest is coupled to several reservoirs, it is important to specify the coarse-grained states not only of the subsystem but also of the reservoirs, in order to proceed with the counting of particles and energy transferred between the reservoirs.  This corresponds to taking two successive quantum measurements, as done in Section~\ref{trFT} and this is sometimes referred to as unraveling the master equation \cite{C93,GMWS01,DRM08}.  Consequently, the master equation rules the density operator of the subsystem conditioned to the numbers of particles and energy that have been transferred between the reservoirs since the initial time.

Many experimental observations are not sensitive to the quantum coherences described by the off-diagonal elements of the density operator.  Under such circumstances, the knowledge of the diagonal elements, i.e., the probabilities of the coarse-grained states are enough for the description of the observations.  In these cases, the time evolution of the probabilities is ruled by a Markovian master equation.  For these stochastic processes, a current fluctuation theorem has been proved using graph analysis \cite{AG06JSM,AG07JSP}.  This theorem is expressed by the symmetry relation (\ref{CFT-bis}) in terms of the cumulant generating function of the fluctuating currents.  Another proof of this theorem is based on the symmetry of the master equation modified to include the counting parameters, as explained here below.

We consider electron transport through an isothermal open subsystem in contact with $r$ reservoirs.  If the circuit is composed of $d$ disconnected subcircuits, the numbers of electrons in these subcircuits are so many conserved quantities.  We suppose that each subcircuit connects at least two reservoirs so that $r\geq 2d$.  In this system, the nonequilibrium conditions are imposed with $p=r-d$ differences of chemical potentials, which define the independent affinities $A_j=\beta(\mu_j-\mu_k)$ with $j=1,2,...,p$ and $k=1,2,...,d$.  These affinities may drive so many independent currents.  We denote by ${\bf n}=\{ n_j\}_{j=1}^p$ the random numbers of electrons transferred from the driving reservoirs $j=1,2,...,p$ to the reference reservoirs $k=1,2,...,d$.  The probabilities ${\bf p}({\bf n})=\{p_{\sigma}({\bf n})\}$ that the subsystem has evolved to some coarse-grained state $\sigma$ while $\bf n$ electrons have been transferred during the time interval $t$ are ruled by the master equation:
\be
\partial_t \, {\bf p}_t({\bf n}) = \hat{\mbox{\helvb L}} \cdot {\bf p}_t({\bf n})
\label{master-eq}
\ee
with the operator
\be
\hat{\mbox{\helvb L}} = \sum_{\rho} \left[ \mbox{\helvb L}_{\rho}^{(+)} \, {\rm e}^{-\frac{\partial}{\partial n_{\rho}}} + \mbox{\helvb L}_{\rho}^{(0)} + \mbox{\helvb L}_{\rho}^{(-)} \, {\rm e}^{+\frac{\partial}{\partial n_{\rho}}} \right]
\label{L-master-eq}
\ee
where $\mbox{\helvb L}_{\rho}^{(\pm)}$ is the matrix with the rates of the transitions $n_{\rho}\to n_{\rho}\pm 1$, while $\mbox{\helvb L}_{\rho}^{(0)}$ is the matrix with the rates of the other transitions and all the loss rates.  Examples of such transport processes in quantum dots have been studied for instance in Refs.~\cite{SKB10,BEG11}. The cumulant generating function is defined as
\be
{\cal Q}(\pmb{\eta}) = \lim_{t\to\infty} \, - \frac{1}{t} \, \ln \left\langle {\rm e}^{-\pmb{\eta}\cdot{\bf n}}\right\rangle_t 
\label{Q-dfn-bis}
\ee
where the statistical average is carried out over the probability distribution ${\bf p}_t({\bf n})$.
As a consequence, the generating function turns out to be given by the eigenvalue problem
\be
\mbox{\helvb L}(\pmb{\eta}) \cdot {\bf v} = -{\cal Q}(\pmb{\eta}) \; {\bf v} 
\label{eigenvalue-eq}
\ee
for the modified matricial operator
\be
\mbox{\helvb L}(\pmb{\eta}) = {\rm e}^{-\pmb{\eta}\cdot{\bf n}}\, \hat{\mbox{\helvb L}} \, {\rm e}^{+\pmb{\eta}\cdot{\bf n}}= \sum_{\rho} \left[ \mbox{\helvb L}_{\rho}^{(+)} \, {\rm e}^{-\eta_{\rho}} + \mbox{\helvb L}_{\rho}^{(0)} + \mbox{\helvb L}_{\rho}^{(-)} \, {\rm e}^{+\eta_{\rho}} \right]  \; .
\label{modified-L}
\ee
Such matricial operators may obey the following symmetry relation
\be
\mbox{\helvb M}^{-1}\cdot \mbox{\helvb L}(\pmb{\eta}) \cdot \mbox{\helvb M}= \mbox{\helvb L}({\bf A}-\pmb{\eta})^{\rm T} 
\ee
where $^{\rm T}$ denotes the matricial transpose and $\mbox{\helvb M}$ is the matrix of the thermal distribution for the subsystem at equilibrium with the $d$ reference reservoirs \cite{K98,BEG11}.
If this symmetry relation holds for the operator (\ref{modified-L}), this property extends to all its eigenvalues and, in particular, to the leading eigenvalue which gives the cumulant generating function.  In this way, the current fluctuation theorem is demonstrated for the subsystem in the nonequilibrium steady state specified by the affinities~$\bf A$:
\be
{\cal Q}(\pmb{\eta}) = {\cal Q}({\bf A}-\pmb{\eta})  \; .
\label{CFT-stoch}
\ee
Accordingly, the results of Section~\ref{Resp} apply here also for the response coefficients.

An equivalent expression of the current fluctuation theorem is given in terms of the probability $P_t({\bf n})$ that $\bf n$ electrons have been transferred during the time interval $t$ between the reservoirs under the stationary conditions~$\bf A$.  Equation (\ref{CFT-stoch}) implies that opposite fluctuations of these numbers have probabilities obeying the symmetry relation:
\be
\frac{P_t({\bf n})}{P_t(-{\bf n})}  \simeq {\rm e}^{{\bf A}\cdot{\bf n}} \qquad \mbox{for} \qquad t\to \infty \; .
\label{CFT-stoch-bis}
\ee
At equilibrium where the affinities are vanishing ${\bf A}={\bf 0}$, the probabilities of opposite fluctuations are equal so that we recover the principle of detailed balancing.  In contrast, a directionality manifests itself out of equilibrium where the non-vanishing affinities introduce a bias between the probabilities.  Soon, one of the probabilities becomes dominant as time increases and the currents tend to flow in one direction so that the mean currents 
\be
{\bf J} = \frac{\partial Q}{\partial\pmb{\eta}}({\bf 0}) = \lim_{t\to\infty} \frac{\langle {\bf n}\rangle_t}{t}
\ee
are no longer vanishing.

\subsection{Thermodynamic entropy production}

A consequence of the current fluctuation theorem is the non-negativity of the entropy production, which is known to be equal to the sum of the affinities multiplied by the mean currents \cite{O31,DD36,P67,C85}. By Jensen's inequality $\langle{\rm e}^{-X}\rangle \geq {\rm e}^{-\langle X\rangle}$ \cite{CT06}, we find that
\be
\left\langle{\rm e}^{-{\bf A}\cdot{\bf n}}\right\rangle_t \geq {\rm e}^{-\langle {\bf A}\cdot{\bf n}\rangle_t}
\ee
Consequently, the entropy production is non negative because
\be
\frac{1}{k_{\rm B}} \, \frac{d_{\rm i}S}{dt} = {\bf A}\cdot{\bf J} = \lim_{t\to\infty} \frac{1}{t} \, {\bf A}\cdot\langle{\bf n}\rangle_t \geq \lim_{t\to\infty} - \frac{1}{t} \, \ln \left\langle{\rm e}^{-{\bf A}\cdot{\bf n}}\right\rangle_t ={\cal Q}({\bf A})={\cal Q}({\bf 0})=0 \; .
\label{entr-prod}
\ee
The last equality is the consequence of the definition (\ref{Q-dfn-bis}) for the cumulant generating function at $\pmb{\eta}=0$.  The second law of thermodynamics can thus be deduced from the current fluctuation theorem.

The entropy production vanishes at equilibrium and is positive out of equilibrium where energy is dissipated. In order to drive a particular current with the other currents, energy should thus be supplied and the second law constitutes a limit to the efficiency of energy transduction.  Powering the process $\gamma$ by the other ones requires that the corresponding mean current is opposite to its  associated affinity: $A_{\gamma} J_{\gamma} <0$.  In this case, a thermodynamic efficiency can be defined and the second law implies that it cannot reach values larger than unity:
\be
0 \leq \eta_{\gamma} \equiv - \frac{A_{\gamma} J_{\gamma}}{\sum_{\alpha\neq\gamma}A_{\alpha} J_{\alpha}} \leq 1 \; .
\ee
Such thermodynamic efficiencies have been considered for molecular motors \cite{GG10} as well as for mass separation by effusion \cite{GA11}.

\subsection{The case of effusion processes}

The current fluctuation theorem has also been established for effusion processes \cite{GA11,CVK06}.  Effusion processes are the classical analogues of electron quantum transport through a constriction, such as the quantum point contact of Fig.~\ref{fig3}a.  In effusion, two gases at different pressures and temperatures are separated by a wall with a pore which is smaller than the mean free path so that the flow of particles across the pore is essentially ballistic. 
The master equation of effusion has been known in classical kinetic theory for a century \cite{K1909,P58}.
The particles of energy $\varepsilon=m{\bf v}^2/2$ are crossing a pore of cross-section area $\sigma$ with the rate:
\be
W_j(\varepsilon) = \frac{\sigma\, n_j}{\sqrt{2\pi m}} \, \beta_j^{3/2} \, \varepsilon \, {\rm e}^{-\beta_j\varepsilon} \qquad\mbox{for}\qquad j={\rm L}, {\rm R}
\label{rates}
\ee
if they come from the left- or the right-hand reservoir at the inverse temperature $\beta_j=(k_{\rm B}T_j)^{-1}$ and the particle density $n_j$ related to the pressure by $P_j=n_j k_{\rm B}T_j$.  These rates satisfy the condition
\be
\frac{W_{\rm L}(\varepsilon)}{W_{\rm R}(\varepsilon)} = {\rm e}^{\varepsilon \, A_E + A_N} 
\label{ratio}
\ee
in terms of the affinities (\ref{A_E}) and (\ref{A_N}).  In effusion, the cumulant generating function of the energy and particle currents is given by 
\be
{\cal Q}(\xi,\eta) = \int_0^{\infty} d\varepsilon\, 
\left[ W_{\rm L}(\varepsilon)  \left( 1 - {\rm e}^{-\varepsilon\xi-\eta}\right) 
+ W_{\rm R}(\varepsilon)  \left( 1 - {\rm e}^{\varepsilon\xi+\eta}\right)\right] \, .
\label{Q-Eff}
\ee
in terms of the transition rates (\ref{rates}) \cite{GA11,CVK06}.  By their property (\ref{ratio}), the generating function obeys the same symmetry relation (\ref{FT-LL}) as in the quantum case.

The generating functions (\ref{Q-LL}) and (\ref{Q-Eff}) of the quantum and classical processes can be compared. At a given energy $\varepsilon$ for $\xi=0$, the generating function has the form (\ref{Q-trinomial}) for the quantum process.
In the limit where $p_{\pm}\ll 1$, it becomes
\be
\tilde{\cal Q}(\eta)= rp_+\, (1-{\rm e}^{-\eta})+ rp_-\, (1-{\rm e}^{\eta})
\label{Q-Poisson}
\ee
which is the classical form appearing in Eq.~(\ref{Q-Eff}) with the identification $rp_{\pm}=W_{\rm L,R}(\varepsilon)$.
The expression (\ref{Q-Poisson}) is characteristic of the combination of two independent Poisson processes, i.e., the particles can be transferred in both directions.

\subsection{Statistics of histories and time reversal}

Another time-reversal symmetry relationship concerns the statistical properties of the histories or paths followed by a system under stroboscopic observations at some sampling time $\Delta t$.  Such observations generate sequences of coarse-grained states such as
\be
\pmb{\omega} = \omega_1\omega_2 \cdots \omega_n
\ee
corresponding to the successive times $t_j=j\, \Delta t$ with $j=1,2,...,n$.  This history or path has a certain probability $P(\pmb{\omega})$ to happen if the system is in the stationary state corresponding to the affinities $\bf A$.  Because of the randomness of the molecular fluctuations, these path probabilities typically decrease exponentially at some rate $h$ that characterizes the temporal disorder in the process.  This characterization applies to stochastic processes as well as chaotic dynamical systems, for which the temporal disorder $h$ is called the Kolmogorov-Sinai or dynamical entropy \cite{ER85,G98,D99,GW93}.  

In nonequilibrium stationary states, the time-reversed path
\be
\pmb{\omega}^{\rm R} = \omega_n  \cdots \omega_2\omega_1
\ee
is expected to happen with a different probability decreasing at the different rate $h^{\rm R}$ now characterizing the temporal disorder of the time-reversed paths \cite{G04JSP}.  The remarkable result is that the difference between the disorders of the time-reversed and typical paths is equal to the thermodynamic entropy production \cite{G04JSP}
\be
\frac{1}{k_{\rm B}} \frac{d_{\rm i}S}{dt} = h^{\rm R} - h \geq 0 
\label{hr-h}
\ee
The second law of thermodynamics is satisfied because this difference is known in mathematics to be a relative entropy, which is always non negative \cite{CT06}.  At equilibrium, detailed balancing holds so that every history and its time reversal are equiprobable, their temporal disorders are equal $h=h^{\rm R}$, and the entropy production vanishes.  This is no longer the case away from equilibrium where the typical paths are more probable than their time reversals.  Consequently, the time-reversal symmetry is broken at the level of the statistical description in terms of the probability distribution of the nonequilibrium stationary state.  In this regard, the entropy production is a measure of the time asymmetry in the temporal disorders of the typical histories and their time reversals.  As a corollary of the second law, the typical histories are more ordered in time than their corresponding time reversals in the sense that $h< h^{\rm R}$ in nonequilibrium stationary states \cite{G07CRP}.

In the case of effusion processes, the temporal disorders of the histories and their time reversals are characterized by
\bea
h&=&\left(\ln\frac{\rm e}{\Delta\varepsilon\Delta t}\right) \int_0^{\infty} d\varepsilon\left[ W_{\rm L}(\varepsilon)+W_{\rm R}(\varepsilon)\right] - \int_0^{\infty} d\varepsilon\left[ W_{\rm L}(\varepsilon)\ln W_{\rm L}(\varepsilon)+W_{\rm R}(\varepsilon)\ln W_{\rm R}(\varepsilon)\right]  + O(\Delta t) \; , \\
h^{\rm R}&=&\left(\ln\frac{\rm e}{\Delta\varepsilon\Delta t}\right) \int_0^{\infty} d\varepsilon\left[ W_{\rm L}(\varepsilon)+W_{\rm R}(\varepsilon)\right] - \int_0^{\infty} d\varepsilon\left[ W_{\rm L}(\varepsilon)\ln W_{\rm R}(\varepsilon)+W_{\rm R}(\varepsilon)\ln W_{\rm L}(\varepsilon)\right]  + O(\Delta t) \; , 
\eea
in terms of the transition rates (\ref{rates}), the sampling time $\Delta t$, and the coarse graining in energy $\Delta\varepsilon$.  The first term of these quantities is a feature of stochastic processes that are continuous in time and energy. For such processes, randomness manifests itself on arbitrarily small time scales $\Delta t$ and energy scales $\Delta\varepsilon$ so that the temporal disorders increase as $\Delta\varepsilon\Delta t\to 0$.

The difference between these two quantities is equal to the entropy production because
\be
 h^{\rm R} - h = \int_0^{\infty} d\varepsilon\left[ W_{\rm L}(\varepsilon)-W_{\rm R}(\varepsilon)\right] \, \ln \frac{W_{\rm L}(\varepsilon)}{W_{\rm R}(\varepsilon)} = A_E \,  J_E + A_N \,  J_N = \frac{1}{k_{\rm B}} \frac{d_{\rm i}S}{dt} \; ,
\label{hr-h-Eff}
\ee
in the limit where $\Delta\varepsilon\Delta t \to 0$.

The validity of the formula (\ref{hr-h}) has also been verified in experiments where the nonequilibrium constraints are imposed by fixing the currents instead of the affinities \cite{AGCGJP07,AGCGJP08}.  

Similar considerations have been developed for quantum systems \cite{G94,CRJMN04}. We notice that quantum mechanics naturally limits the randomness on the scale where $\Delta\varepsilon\Delta t = 2\pi \hbar$, as explained in Ref. \cite{G94}. 

\section{Transport in electronic circuits}
\label{Electron}

Several types of electronic circuits are considered such as single-electron transistors, quantum dots with resonant levels, or quantum dots capacitively coupled with a quantum point contact \cite{FGB09,NB09,BB00}.

\subsection{Quantum dot with one resonant level}

A quantum dot is a kind of artificial atom with quantized electronic levels.  Because of the tunneling to the reservoirs, these levels are broadened into resonances characterized by a lifetime.  We suppose that the transport process involves a single resonant level, which may thus be either empty or occupied.  In this case, the process is ruled by the master equation (\ref{master-eq}) with a modified operator (\ref{modified-L}) given by
\be
{\mbox{\helvb L}}(\eta) =
\left(
\begin{array}{cc}
-a_{\rm L}-a_{\rm R} &  b_{\rm L}{\rm e}^{+\eta}+b_{\rm R}  \\
a_{\rm L}{\rm e}^{-\eta}+a_{\rm R} & -b_{\rm L}-b_{\rm R} \\
\end{array}
\right)
\label{modified-L-QD}
\ee
in terms of the following charging and discharging rates:
\be
a_{\rho} = \Gamma_{\rho} \, f_{\rho} \qquad\mbox{and}\qquad
b_{\rho} = \Gamma_{\rho} \, (1-f_{\rho}) \qquad\mbox{for}\qquad \rho={\rm L}, {\rm R} \; .
\ee
The rate constants $\Gamma_{\rho}$ are
proportional to the square of the interaction parameters between
the quantum dot and the reservoir $\rho$, as well as to the density of states of 
the reservoir $\rho$.  The Fermi-Dirac distributions are given by
\be
f_{\rho} = \frac{1}{1+{\rm e}^{\beta(\varepsilon_0+\Delta U_0 -\mu_{\rho})}}  \qquad\mbox{for}\qquad \rho={\rm L},{\rm R}
\ee
where $\varepsilon_0$ is the bare energy of the dot level, $\Delta U_0$
is the electrostatic charging energy, 
and $\mu_{\rho}=eV_{\rho}$ are the chemical potentials of the reservoirs \cite{FGB09,NB09,SB11}.

The cumulant generating function is given by the eigenvalue problem (\ref{eigenvalue-eq}) for the matrix (\ref{modified-L-QD}) as
\be
{\cal Q}(\eta) =  \frac{1}{2} \left[ 
a_{\rm L}+a_{\rm R} +b_{\rm L}+b_{\rm R} -
\sqrt{\left(a_{\rm L}+a_{\rm R} -b_{\rm L}-b_{\rm R}\right)^2
+ 4 \left( a_{\rm L}{\rm e}^{-\eta}+a_{\rm R}\right)\left(b_{\rm L}{\rm e}^{+\eta}+b_{\rm R}\right)}\right]
\label{Q-QD}
\ee
This function obeys the symmetry relation ${\cal Q}(\eta)= {\cal Q}(A-\eta)$ whereupon the current fluctuation theorem (\ref{CFT-stoch-bis}) is satisfied with the affinity \cite{AG06JSM}
\be
A = \ln \frac{a_{\rm L} \, b_{\rm R}}{a_{\rm R} \, b_{\rm L}} = \beta(\mu_{\rm L}-\mu_{\rm R}) = \beta e V \; .
\ee

\subsection{Capacitively coupled circuits}

In order to perform the full counting statistics of electron transport in quantum dots, the main circuit should be monitored by a secondary circuit which is made of a quantum point contact (QPC) \cite{GLSSISEDG06,GLSSIEDG09,FHTH06}.  Both circuits are capacitively coupled so that the current in the QPC is sensitive to the occupancy of the QDs.  This latter modulates the QPC current by Coulomb repulsion.  In this way, the quantum jumps in the occupancy of the QDs can be experimentally observed.  Typically, the electric current in the QPC is seven or eight orders of magnitude larger than the current in the circuit with the QDs (see Table~\ref{Table1}). From the viewpoint of quantum measurement, the QPC plays the role of a measuring device and the QDs the small observed quantum system. Because of the large ratio between their currents, the measuring device is essentially in a classical regime.  Although the ratio of the noise to the current is negligible in this regime, the noise gets larger as the current increases.  As a consequence, the QDs themselves are subjected to an important noise, which affects their kinetics. In particular, the charging and discharging rates of the QDs are modified by the presence of the large current in the QPC.  Therefore, the Coulomb interaction between both circuits also causes a back action of the QPC onto the QDs.

\begin{figure}[h]
\begin{center}
\includegraphics[scale=0.3]{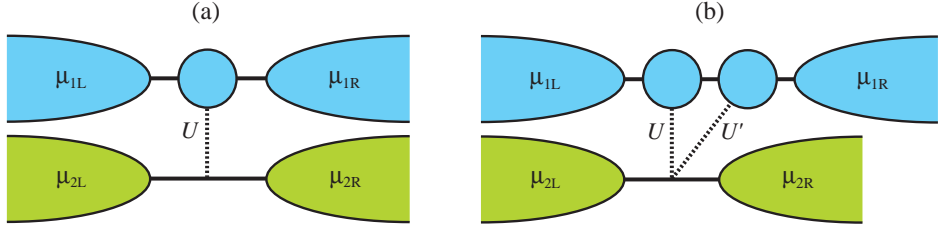}
\caption{(a) Schematic representation of a single quantum dot (QD) monitored by a quantum point contact (QPC) to which it is capacitively coupled by the Coulomb interaction $U$, as in the circuit of Ref.~\cite{GLSSISEDG06}. (b) Schematic representation of two QDs monitored by a QPC asymmetrically positioned with respect to the QDs in order to get different capacitive couplings of the QPC to the two QDs, as in the circuits of Refs.~\cite{FHTH06,UGMSFS10,KRBMGUIE12}.  The Coulomb repulsion $U$ is larger than $U'$, which allows the bidirectional counting of electrons in the circuit with the two QDs.}
\label{fig5}
\end{center}
\end{figure}

\begin{table}
\begin{centering}
\begin{tabular}{|c|c|c|}
\hline 
  & \ single QD \cite{GLSSISEDG06} \ & \ double QD \cite{FHTH06} \ \tabularnewline
\hline 
$T$ & $350$ mK & $130$ mK \tabularnewline
  &  &  \tabularnewline
$I_{\rm QD}$ & $1.3\times 10^{-16}$ A & $6.5\times 10^{-17}$ A \tabularnewline
$V_{\rm QD}$ & $2.7$ mV & $0.3$ mV \tabularnewline
$\Pi_{\rm QD}$ & $3.4\times 10^{-19}$ W & $2.0\times 10^{-20}$ W \tabularnewline
$J_{\rm QD}$ & $792$ Hz & $406$ Hz \tabularnewline
$A_{\rm QD}$ & $90$ & $27$ \tabularnewline
  &  &  \tabularnewline
$I_{\rm QPC}$ & $4.5\times 10^{-9}$ A & $1.2\times 10^{-8}$ A \tabularnewline
$V_{\rm QPC}$ & $0.5$ mV & $0.8$ mV \tabularnewline
$\Pi_{\rm QPC}$ & $2.3\times 10^{-12}$ W & $9.6\times 10^{-12}$ W \tabularnewline
$J_{\rm QPC}$ & $2.8\times 10^{10}$ Hz & $7.5\times 10^{10}$ Hz \tabularnewline
$A_{\rm QPC}$ & $17$ & $71$ \tabularnewline
  &  &  \tabularnewline
$\frac{I_{\rm QPC}}{I_{\rm QD}}$ & $3.5\times 10^{7}$ & $1.8\times 10^{8}$ \tabularnewline
\hline
\end{tabular}
\par\end{centering}
\caption{Electronic temperature $T$, mean electric currents $I_{\alpha}$, voltages $V_{\alpha}$, dissipated powers $\Pi_{\alpha}=V_{\alpha} I_{\alpha}$, mean electron currents $J_{\alpha}= I_{\alpha}/\vert e\vert$, affinities $A_{\alpha}=\vert e\vert V_{\alpha}/(k_{\rm B}T)$ in the quantum dot ($\alpha={\rm QD}$) and the quantum point contact ($\alpha={\rm QPC}$), as well as the ratio of QPC to QD currents, for the experiments reported in Ref.~\cite{GLSSISEDG06} with a single QD and Ref.~\cite{FHTH06} with a double QD.  According to Eq.~(\ref{entr-prod}), the thermodynamic entropy production in the whole system is equal to $d_{\rm i}S/dt =k_{\rm B}\sum_{\alpha}A_{\alpha} J_{\alpha}=\sum_{\alpha}\Pi_{\alpha}/T$, which is dominated by the energy dissipation in the QPC under these experimental conditions.}
\label{Table1} 
\end{table}

Several types of circuits have been considered.  
In Ref.~\cite{GLSSISEDG06}, a single QD with a strong bias voltage is monitored by a QPC.
The QD is successively empty or occupied so that the QPC current is observed to jump between both corresponding values. This setup allows to carry out the full counting statistics of the occupancy of the QD.  Since the bias voltage is large $\mu_{1{\rm L}}\gg\mu_{1{\rm R}}$, the transition rates of the backward fluctuations can be assumed to be negligible.  Under these conditions, the statistics of the fluctuating current can be inferred from the statistics of the occupancy. However, the backward fluctuations of the current are also negligible in this fully irreversible regime.  Therefore, the current fluctuation theorem cannot be tested for lack of backward fluctuations.

In order to overcome such limitations, a setup with two QDs has been considered in Ref.~\cite{FHTH06}.  The two QDs are positioned at different distances from the QPC so that different Coulomb interactions $U$ and $U'$ exist between the QPC and both QDs, as shown in Fig.~\ref{fig5}b.  If an electron travels from the left- to the right-hand QD, the Coulomb repulsion is successively larger and smaller, or vice versa for the reversed motion of an electron.  Therefore, this setup is suitable for bidirectional counting statistics.  Remarkably, the symmetry relation $P_t(n) = P_t(-n) \exp(\tilde A_{\rm QD} \, n)$ is observed but, depending on the experimental conditions \cite{FHTH06,UGMSFS10,KRBMGUIE12}, the effective affinity $\tilde A_{\rm QD}$ may significantly differ from the thermodynamic affinity $A_{\rm QD}=\vert e\vert V_{\rm QD}/(k_{\rm B}T)$ driving the current in the QD.  In Refs.~\cite{FHTH06,UGMSFS10}, the gate voltages of the QDs are tuned so that transitions occur between charge states where the QDs are singly or doubly occupied and
the effective affinity $\tilde A_{\rm QD}\simeq 3$ differs by one order of magnitude from the thermodynamic affinity $A_{\rm QD} = 27$ (see Table~\ref{Table1}).
In Ref.~\cite{KRBMGUIE12}, the experiment has been carried out closer to equilibrium with an optimized sample design and gate voltages tuned so that transitions occur between charge states where the QDs are empty or singly occupied.  For the bias voltage $V_{\rm QD}=20 \;\mu{\rm V}$ and the electronic temperature $T=330$ mK, the thermodynamic affinity is equal to $A_{\rm QD} = 0.7$ while the effective affinity is observed to be $\tilde A_{\rm QD} = 0.5$.  The differences observed in the experiments reported in Refs.~\cite{FHTH06,UGMSFS10,KRBMGUIE12} arise because the QPC may induce back-action effects onto the dynamics of the QDs under some conditions, while the limited bandwidth of the charge detection also affects the counting statistics.

\begin{figure}[h]
\begin{center}
\includegraphics[scale=0.3]{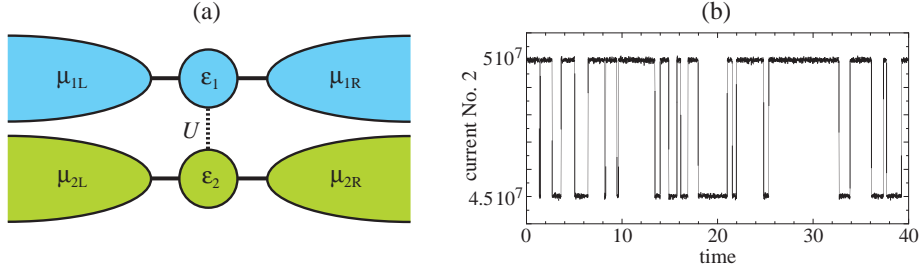}
\caption{(a) Schematic representation of two quantum dots in parallel.  Each quantum dot is coupled
to two reservoirs of electrons.  Moreover, both quantum dots influence each other by the Coulomb electrostatic interaction $U$. (b) Simulation with Gillespie's algorithm \cite{G76} of the detector current in circuit No.\,2 measuring the occupancy of the QD No.\,1.  The parameter values are given by $\beta\varepsilon_1=0$, $\beta\varepsilon_2 = 35$, $\beta U = 32.8$, 
$\beta\mu_{1{\rm L}} = 25$,
$\beta\mu_{1{\rm R}} = 0$,
$\Gamma_{1{\rm L}}= \Gamma_{1{\rm R}} =\bar{\Gamma}_{1{\rm L}}= \bar{\Gamma}_{1{\rm R}} = 1$,
$\beta\mu_{2{\rm L}} = 70$,
$\beta\mu_{2{\rm R}} = 0$, and
$\Gamma_{2{\rm L}}= \Gamma_{2{\rm R}} = \bar{\Gamma}_{2{\rm L}}= \bar{\Gamma}_{2{\rm R}} = 10^8$.  
The affinities of both circuits are $A_1=25$ and $A_2=70$.
The effective affinity of the circuit No.\,1 is $\tilde{A}_1=1.17$.  The mean value of the QD current is $ J_1 \simeq 0.17$ electrons per unit time.  The mean value of the secondary current is $ J_2 \simeq 4.8\times10^7$ electrons per unit time.  The QD is empty (respectively occupied) when the secondary current takes the value $5\times 10^7$ (respectively $4.5\times 10^7$).  Adapted from Ref.~\cite{BEG11}.}
\label{fig6}
\end{center}
\end{figure}

In order to understand the back-action effects, the capacitive coupling between the QDs and the QPC should be taken into account. For this purpose, the simple model of Fig.~\ref{fig6}a with two capacitively coupled QDs in two parallel circuits may be considered \cite{SLSB10,SKB10,BEG11}.  The system is composed of $d=2$ disconnected circuits and $r=4$ reservoirs, so that the nonequilibrium steady states are specified with $p=r-d=2$ affinities:
\be
A_1 = \beta(\mu_{1{\rm L}}-\mu_{1{\rm R}}) \qquad\mbox{and}\qquad A_2 = \beta(\mu_{2{\rm L}}-\mu_{2{\rm R}}) \; .
\label{A1-A2}
\ee
The system can be described in terms of the probabilities $p_{\nu_1\nu_2}(n_1,n_2)$ that each quantum dot ($\alpha=1$ or $2$) is occupied by $\nu_{\alpha}=0$ or $1$ electron, while $n_{\alpha}$ electrons have been transferred from the left- to the right-hand reservoirs in the circuit $\alpha$ during the time interval $t$.
The stochastic process is ruled by the master equation (\ref{master-eq}) with the operator (\ref{L-master-eq}).  For $\alpha=1,2$ and $\rho={\rm L},{\rm R}$, the charging and discharging rates are given by
\bea
\mbox{either}\qquad&& a_{\alpha\rho} = \Gamma_{\alpha\rho}  f_{\alpha\rho} \qquad\mbox{and}\qquad 
b_{\alpha\rho}  = \Gamma_{\alpha\rho}  (1-f_{\alpha\rho} ) \qquad\mbox{with}\qquad  
f_{\alpha\rho}  = \frac{1}{{\rm e}^{\beta(\varepsilon_{\alpha}-\mu_{\alpha\rho})}+1} \; ,
\label{a-b} \\
\mbox{or}\qquad 
&& \bar{a}_{\alpha\rho} = \bar{\Gamma}_{\alpha\rho}  \bar{f}_{\alpha\rho} \qquad\mbox{and}\qquad 
\bar{b}_{\alpha\rho}  = \bar{\Gamma}_{\alpha\rho}  (1-\bar{f}_{\alpha\rho} ) \qquad\mbox{with}\qquad  
\bar{f}_{\alpha\rho}  = \frac{1}{{\rm e}^{\beta(\varepsilon_{\alpha}+U-\mu_{\alpha\rho})}+1} \; ,
\label{bar-a-b}
\eea
if the other quantum dot is either empty or occupied.  The effect of the tunneling barrier capacitances is neglected in this simplified model \cite{SKB10,BEG11}.  We notice that their effect has been analyzed in Ref.~\cite{SLSB10}.

\begin{figure}[h]
\centerline{\includegraphics[scale=0.35]{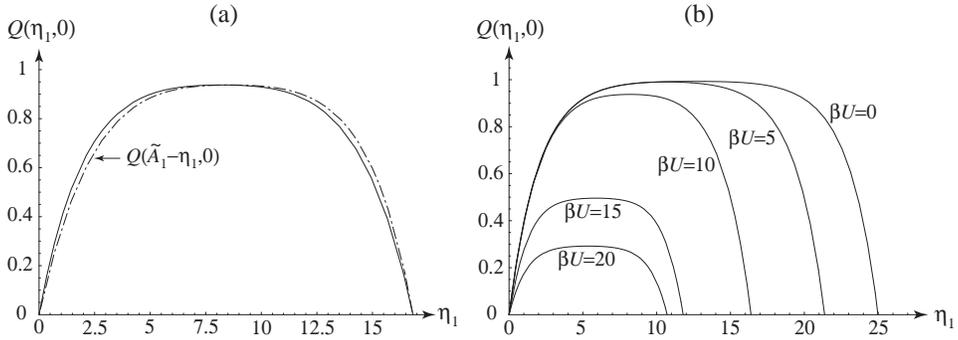}}
\caption{The cumulant generating function ${\cal Q}(\eta_1,\eta_2=0)$ for the circuit of Fig.~\ref{fig6}a   versus the counting parameter $\eta_1$:
(a)~together with the symmetric function with respect to the effective affinity $\tilde{A}_1=16.8356$ (dotted-dashed line) for
$\beta U=10$ and $\Gamma_{2{\rm L}}=\Gamma_{2{\rm R}}= \bar{\Gamma}_{2{\rm L}}= \bar{\Gamma}_{2{\rm R}}=2$;
(b)~for different values of the electrostatic coupling parameter $\beta U$ and 
$\Gamma_{2}\equiv\Gamma_{2{\rm L}}=\Gamma_{2{\rm R}}=\bar{\Gamma}_{2{\rm L}}= \bar{\Gamma}_{2{\rm R}}=100$.
In both cases, the other parameters take the following values: 
$\beta\varepsilon_1=10$, $\beta\varepsilon_2=35$, 
$\beta\mu_{1{\rm L}}=25$, $\beta\mu_{1{\rm R}}=0$, 
$\beta\mu_{2{\rm L}}=70$, $\beta\mu_{2{\rm R}}=0$, and
$\Gamma_{1}\equiv\Gamma_{1{\rm L}}=\Gamma_{1{\rm R}}=\bar{\Gamma}_{1{\rm L}}= \bar{\Gamma}_{1{\rm R}} =1$. In these plots, the effective affinity is given by the second root of the function at $\eta_1=\tilde A_1 \neq 0$.  The thermodynamic affinity of the secondary circuit is $A_2=\beta(\mu_{2{\rm L}}-\mu_{2{\rm R}})=70$. Adapted from Ref.~\cite{BEG11}.}
\label{fig7}
\end{figure}

Now, the cumulant generating function (\ref{Q-dfn-bis}) is obtained by solving the eigenvalue problem (\ref{eigenvalue-eq}) and it obeys the current fluctuation theorem (\ref{CFT-stoch}) for the two currents flowing across the parallel circuits:
\be
{\cal Q}(\eta_1,\eta_2) = {\cal Q}(A_1-\eta_1,A_2-\eta_2)
\ee
with respect to the two affinities (\ref{A1-A2}).  However, this result does not imply that the symmetry relation holds for the single current flowing in one of the circuits.  The generating function of this single current is given by ${\cal Q}(\eta_1,0)$ and the symmetry does not hold in general, 
${\cal Q}(\eta_1,0) \neq {\cal Q}(A_1-\eta_1,0)$, as shown in the counterexample of Fig.~\ref{fig7}a.  

Nevertheless, under some circumstances, the single-current fluctuation theorem
\be
{\cal Q}(\eta_1,0) = {\cal Q}(\tilde A_1-\eta_1,0)
\label{SCFT}
\ee
holds with respect to an effective affinity $\tilde A_1$, which can be calculated.

This symmetry holds with the effective affinity equal to the thermodynamic affinity, $\tilde A_1=A_1$, in a few cases: (1) if the other circuit is at equilibrium, i.e., if $A_2=0$; (2) if there is no Coulomb interaction between both circuits, $U=0$, in which case the two circuits are independent of each other; (3) in
the limit where the Coulomb interaction is very large, $U=\infty$. Indeed, the probability that both quantum dots are occupied is vanishing in this limit, which reduces the Markov process to three states and implies the symmetry (\ref{SCFT}) \cite{BEG11}.

The single-current fluctuation theorem (\ref{SCFT}) also holds in the limit where the rate constants of one circuit are much larger than in the other circuit: 
\be
\Gamma_{\rm 2\rho}, \bar{\Gamma}_{\rm 2\rho} \gg \Gamma_{\rm 1\rho}, \bar{\Gamma}_{\rm 1\rho} \qquad\mbox{for}\qquad \rho={\rm L},{\rm R}
\label{conditions}
\ee
as in the experiments of Refs.~\cite{GLSSISEDG06,FHTH06} where the ratio between the QPC and QD currents is many orders of magnitude larger than unity (see Table~\ref{Table1}).  In the system of Fig.~\ref{fig6}a, the fast QD in the secondary circuit rapidly jumps between its two states $\nu_2=0$ and $\nu_2=1$ so that the charging and discharging rates of the slow QD are effectively averaged over both states.  The conditional probabilities $P_{\nu_2\vert\nu_1}$ that the second QD has the occupancy $\nu_2$ provided that the first QD is in the state $\nu_1$ are given by
\bea
&& P_{0\vert 0} = \frac{b_{2{\rm L}} + b_{2{\rm R}}}{a_{2{\rm L}} + a_{2{\rm R}}+b_{2{\rm L}} + b_{2{\rm R}}} \; , \qquad
P_{1\vert 0} = \frac{a_{2{\rm L}} + a_{2{\rm R}}}{a_{2{\rm L}} + a_{2{\rm R}}+b_{2{\rm L}} + b_{2{\rm R}}} \; , \\
&& P_{0\vert 1} = \frac{\bar{b}_{2{\rm L}} + \bar{b}_{2{\rm R}}}{\bar{a}_{2{\rm L}} + \bar{a}_{2{\rm R}}+\bar{b}_{2{\rm L}} + \bar{b}_{2{\rm R}}} \; , 
\qquad
P_{1\vert 1} = \frac{\bar{a}_{2{\rm L}} + \bar{a}_{2{\rm R}}}{\bar{a}_{2{\rm L}} + \bar{a}_{2{\rm R}}+\bar{b}_{2{\rm L}} + \bar{b}_{2{\rm R}}} \; .
\eea
Therefore, the effective charging and discharging rates of the first QD are given by the following averages:
\be
\tilde a_{\rho} = a_{1\rho} \, P_{0\vert 0} + \bar{a}_{1\rho} \, P_{1\vert 0}  \qquad\mbox{and}\qquad
\tilde b_{\rho} = b_{1\rho} \, P_{0\vert 1} + \bar{b}_{1\rho} \, P_{1\vert 1}  \qquad\mbox{with}\qquad \rho={\rm L},{\rm R} \; .
\label{eff-rates}
\ee
Under the conditions (\ref{conditions}), the Markovian process is similar to the one ruled by the operator (\ref{modified-L-QD}) but with these effective charging and discharging rates.  Accordingly, 
the cumulant generating function has the form (\ref{Q-QD}) with the rates (\ref{eff-rates}) so that
the single-current fluctuation theorem (\ref{SCFT}) is indeed satisfied with the effective affinity:
\be
\tilde A_1 \equiv \ln \frac{\tilde a_{\rm L} \tilde b_{\rm R}}{\tilde a_{\rm R} \tilde b_{\rm L}} \; .
\label{eff_A_1}
\ee

Figure~\ref{fig7}b shows not only that the single-current generating function ${\cal Q}(\eta_1,0)$ 
is nearly symmetric already if $\Gamma_2/\Gamma_1=100$, but also that the symmetry holds
with respect to an effective affinity $\tilde A_1$ which varies with the electrostatic interaction $\beta U$.
If the Coulomb interaction vanishes $\beta U=0$, both circuits are decoupled so that the thermodynamic
affinity $A_1=25$ is recovered.  However, the effective affinity decreases if the electrostatic interaction increases till $\beta U=20$, as seen in Fig.~\ref{fig7}b.  For larger interaction $\beta U$, the effective affinity (\ref{eff_A_1}) recovers its thermodynamic value in the limit $U\to\infty$, as explained here above \cite{BEG11}. 

The conclusion is that the secondary circuit monitoring the quantum jump dynamics in quantum dots
may have important back-action effects on the full counting statistics.  In general, the current fluctuation theorem holds at the fundamental level of description for all the currents interacting in the system.
In this regard, the secondary circuit performing the measurement of the quantum state in the first circuit
is part of the total system and cannot be separated in general.
Nevertheless, the symmetry relation of the fluctuation theorem is still observed to hold
for a single fluctuating current, but with an effective affinity that may differ from the thermodynamic value due to the back-action effect of the monitoring circuit onto the observed circuit.
The effective affinity (\ref{eff_A_1}) may drop down to a value more than one order of magnitude lower than the thermodynamic affinity \cite{BEG11}.  Similar conclusions have been reached with related approaches where the averages (\ref{eff-rates}) of the transition rates have been carried out in the framework of the $P(E)$-theory \cite{UGMSFS10,GUMS11,IN92}.

A general remark is that quantum measurement implies the dissipation of energy.  In order to observe the quantum jumps of the QD with a short enough time resolution, the current and the dissipated power in the monitoring circuit should be higher than in the QD. If the QD state is monitored with a sampling time $\Delta t$, the secondary circuit playing the role of the detector should have a current $J_2 \gtrsim (\Delta t)^{-1}$.  Hence, the dissipated power should be bounded by $\Pi_2=k_{\rm B}T A_2J_2 \gtrsim k_{\rm B}T A_2(\Delta t)^{-1}$.  The higher the time resolution, the higher the dissipation rate. Therefore,  the experimental conditions are variously  influenced by the bandwidth of charge detection.

\subsection{Coherent quantum conductor}

The current fluctuation theorem also applies to transport in coherent quantum conductors such as Aharonov-Bohm rings in a magnetic field \cite{I97,FGB09,NB09}. We suppose that the circuit is connected to two reservoirs so that there is a single affinity.  In this case, the results of Section~\ref{Resp} are the following. The particle current is expanded in powers of the affinity according to Eq.~(\ref{dfn-currents}):
\be
J(A;{\cal B}) = L({\cal B}) \, A + \frac{1}{2} \, M({\cal B}) \, A^2 + \cdots
\label{J-A}
\ee
Similar expansions can be introduced for the diffusivity (\ref{D}) and the third cumulant (\ref{C}):
\bea
&& D(A;{\cal B}) = D_0({\cal B}) + D_1({\cal B}) \, A  + \cdots \; , \label{D-A}\\
&& C(A;{\cal B}) = C_0({\cal B}) + \cdots \; . \label{C-A}
\eea
As shown in Section~\ref{Resp}, all these coefficients are interconnected by the following relations.
For linear response and the second cumulant, Eqs.~(\ref{Casimir})-(\ref{FD-thm}) give
\be
L({\cal B}) = L(-{\cal B}) = D_0({\cal B}) \; .
\ee
For the third cumulant and the second response coefficient, Eqs.~(\ref{C0-D1})-(\ref{M-D1-C0}) give
\bea
&& C_0({\cal B})=-C_0(-{\cal B}) = 2 \, \left[ D_1({\cal B}) -D_1(-{\cal B})  \right] \; , \label{C0-asym}\\
&& M({\cal B})+M(-{\cal B}) = 2 \, \left[ D_1({\cal B}) +D_1(-{\cal B})  \right] \; , \\
&& M({\cal B}) = 2 \, D_1({\cal B}) - \frac{1}{3}\, C_0({\cal B}) \; .
\eea
Similar relations hold at higher orders \cite{AGMT09}.

In order to compare with usual quantities, we notice that the electric current is equal to $I(V;{\cal B})=eJ(A;{\cal B})$ and the affinity $A$ is related to the voltage $V$ by
\be
A = \frac{eV}{k_{\rm B}T} \; .
\ee
Moreover, the noise power (\ref{noise_power}) at zero frequency $S=S(\omega=0)$ is proportional to the diffusivity according to Eq.~(\ref{S-D}). It is standard to expand the electric current and the noise power as follows:
\bea
&& I(V;{\cal B}) = G_1({\cal B}) \, V + \frac{1}{2} \, G_2({\cal B}) \, V^2  + \cdots \; , \\
&& S(V;{\cal B}) = S_0({\cal B}) + S_1({\cal B}) \, V + \cdots \; .
\eea
Comparing with the expansions (\ref{J-A}) and (\ref{D-A}), we get
\bea
&& e^2 \, L = k_{\rm B}T \, G_1 \; , \qquad e^3 \, M = (k_{\rm B}T)^2 \, G_2 \; ,  \quad\dots \\
&& 4 e^2 \, D_0 = S_0 \; , \qquad\quad 4 e^3 \, D_1 = k_{\rm B}T \, S_1 \; , \quad\dots 
\eea

Accordingly, we recover the Johnson-Nyquist and Casimir-Onsager relations \cite{BB00}
\be
S_0({\cal B}) = 4 \, k_{\rm B}T \, G_1({\cal B}) \qquad\mbox{and}\qquad  G_1({\cal B}) = G_1(-{\cal B}) \; .
\ee
For the nonlinear response properties, the following relations can be deduced
\bea
&& S_1({\cal B}) + S_1(-{\cal B}) = 2\, k_{\rm B}T \, \left[ G_2({\cal B}) +G_2(-{\cal B}) \right] \; , \label{S1+}\\
&& S_1({\cal B}) - S_1(-{\cal B}) = 6\, k_{\rm B}T \, \left[ G_2({\cal B}) -G_2(-{\cal B}) \right] \; , \label{S1-}
\eea
at the level of the third cumulant $C_0({\cal B})$,
as well as higher-order relations.

The expression (\ref{S1+}) is only the consequence of the second equality in Eq.~(\ref{Q=0}), ${\cal Q}(A,A;{\cal B})=0$, which has been called the `global detailed balancing condition' and
which is weaker than the current fluctuation theorem itself \cite{FB08}.
The expression (\ref{S1-}) is the consequence of the current fluctuation theorem 
and, thus, of the underlying microreversibility in the sense that
$\hat\Theta \hat H({\cal B})\hat\Theta^{-1}= \hat H(-{\cal B})$.  
We notice that the quantity (\ref{S1-}) is proportional to the magnetic asymmetry given by the third cumulant (\ref{C0-asym}).

The symmetry relations (\ref{S1+}) and (\ref{S1-}) have been tested in recent experiments 
on a circuit with an Aharonov-Bohm ring in an external magnetic field \cite{NYHCKOLESUG10,NYHCKOLESUG11}.  The circuit is fabricated by local oxidation on a GaAs/AlGaAs 2DEG and the ring has a diameter of about 500 nm.  The electronic temperature $T=125$ mK is determined with the Johnson-Nyquist fluctuation-dissipation relation.  The current is observed as a function of the applied voltage $V$, the gate voltage $V_g$, and the external magnetic field $\cal B$.  The noise power is measured with the cross-correlation technique between the signals from two sets of resonant circuit and amplifier, which are external to the circuit.  Beyond linear response, the coefficients $S_1$ and $G_2$ have been measured and they have complex dependences on the magnetic field and the gate voltage.  Remarkably, the proportionalities predicted by Eqs.~(\ref{S1+}) and (\ref{S1-}) are confirmed. The proportionality constants are observed to be $12.00\pm 1.96$ instead of $2$ in Eq.~(\ref{S1+}) and $9.66\pm 1.32$ instead of $6$ in Eq.~(\ref{S1-}) \cite{NYHCKOLESUG11}.  The antisymmetric relation (\ref{S1-}) is in better agreement with theory than the symmetric relation (\ref{S1+}).  The reason for the considerable deviation in the case of the symmetric relation (\ref{S1+}) is not explained in Refs.~\cite{NYHCKOLESUG10,NYHCKOLESUG11}.  The amplifiers used to measure the noise power may have back-action effects.  

The fundamental questions associated with these issues deserve further experiments to understand the origins of the quantitative deviations and to test predictions at the higher orders if the concerned quantities are experimentally accessible.

\section{Conclusions}
\label{Conclusions}

This chapter has been devoted to recent advances in the nonequilibrium statistical mechanics of quantum systems.  For more than a century, statistical mechanics has offered the conceptual framework to formulate the various statistical aspects of mechanics.  These aspects play a prominent role in quantum mechanics, which is an initial-condition theory as well as classical mechanics.
In such theories, the non-reproducibility of random phenomena is naturally explained by the
arbitrariness of initial conditions in due respect to the principle of causality.
This is the case in open systems under nonequilibrium conditions sustaining currents of energy and particles between reservoirs at different temperatures and chemical potentials.
After transients, the system evolves towards a nonequilibrium steady state which describes the different properties such as the mean currents and their statistical cumulants.  The set of all these
cumulants characterize the full counting statistics and, thus, the large-deviation properties of the fluctuating currents, which have been the focus of recent advances in nonequilibrium statistical mechanics.

These advances find their origins in earlier work on chaotic dynamical systems modeling transport properties such as diffusion, viscosity or heat conductivity \cite{PH88,GN90,GB95,DG95,GD95,vBDPD97,G98,D99,GCGD01,K07,EM08}. 
In this context, large-deviation properties have been studied for physical quantities fluctuating in time.
Previously, the large-deviation properties were mainly considered in equilibrium statistical mechanics
to infer thermodynamics from the spatial fluctuations of observables in the large-size limit.
In dynamical systems theory, large-deviation properties have been introduced in close relation with the concepts of Lyapunov exponent, Kolmogorov-Sinai entropy per unit time, and fractal dimensions \cite{ER85}.  Furthermore, relationships have been established between these quantities characterizing dynamical chaos and the transport properties \cite{PH88,GN90,GB95,DG95,GD95,vBDPD97,G98,D99,GCGD01,K07,EM08}.  In this context, the consideration of time-reversal symmetry for nonequilibrium steady states has opened new perspectives for a fundamental understanding of nonequilibrium systems at small scales where the microscopic degrees of freedom manifest themselves as fluctuations.  Several types of time-reversal symmetry relations have been established \cite{J97,C98,C99,J00,ECM93,GC95,G96,K98,LS99,M99,ES02,vZC03,ADEN10,MMO11,J11,MN03,G04JSP,G05,KPV07}.  On the one hand, so-called fluctuation theorems have been proved which compare the probabilities of opposite fluctuations for nonequilibrium quantities of interest. On the other hand, the thermodynamic entropy production has been shown to result from the breaking of time-reversal symmetry at the statistical level of description.  These advances have turned out to play a fundamental role in the understanding of many nonequilibrium nanosystems such as molecular machines.

More recently, time-reversal symmetry relations have been extended to quantum systems and, in particular, to electron quantum transport in semiconducting nanodevices \cite{CRJMN04,K00,T00,M03,TN05,EHM07,HEM07,SU08,AG08PRL,AGMT09,SD07,TH07,TLH07,CHT11,GK08,EHM09}. Several approaches have been followed.

Functional relations have been established which provide a unifying framework to
deduce many results such as the Casimir-Onsager reciprocity relations, the Kubo formula of linear response theory,  the quantum version of Jarzynski's nonequilibrium work equality, as well as their generalization beyond linear response or to novel physical contexts \cite{AG08PRL}.

Moreover, a theoretical approach has been developed
for the transport of energy and particles across a driven open quantum system between reservoirs under nonequilibrium conditions.  In quantum mechanics, the issue of measurement
is always subtle because of the non-separability of a system into its different parts.
Here, quantum measurements at the initial and final times are required in order to determine
the amounts of energy and particles transferred between the reservoirs during the lapse of time when the system is driven by time-dependent forces.
Using the time-reversal symmetry of the microscopic dynamics or its extension in the
presence of an external magnetic field, symmetry relations have been obtained which are
quantum versions of the transitory current fluctuation theorem.

In the long-time limit, this theorem leads to the stationary current fluctuation theorem.
An essential aspect is that, for nonequilibrium stationary conditions, the symmetry relation 
should hold with respect to the affinities or thermodynamic forces due to the {\it differences}
of temperatures or chemical potentials between the reservoirs.  Remarkably, a proposition
guarantees that the cumulant generating function of the fluctuating currents indeed
depends only on the differences between the counting parameters associated with
the reservoirs between which the thermodynamic forces are maintained \cite{AGMT09}.
Thanks to this proposition enunciated in Section~\ref{stFT}, the stationary current fluctuation theorem
can be established for all the currents flowing across open quantum systems.

Close to equilibrium, known results of linear response theory such as the Casimir-Onsager reciprocity relations and fluctuation-dissipation formulae can be deduced from the stationary current fluctuation theorem. The fact is that this theorem also holds arbitrarily far away from equilibrium, allowing the
generalization of these results to nonlinear response properties \cite{AG04,AG06JSM,AG07JSM,AGMT09}.

These advances concern in particular electron quantum transport in mesoscopic semiconducting circuits. At low temperature, electrons have ballistic motions on the size of such circuits, which form
electronic waveguides.  Many transport properties of mesoscopic conductors can thus be understood
by supposing that the electrons are independent of each other and undergo scattering
in the effective potential formed by the semiconducting circuit.
In this approximation where electron-electron and electron-phonon interactions are
neglected,  the cumulant generating function of the aforementioned theorem can be expressed
with the Levitov-Lesovik formula in terms of the scattering matrix of the independent electrons on the obstacles of the circuit.  Therefore, this generating function has the time-reversal symmetry of the fluctuation theorem with respect to the affinities driving the currents across the circuit.
This fundamental result has many consequences, in particular, concerning the full counting statistics of electrons and the relationships between conductance and noise power beyond linear response.  Recently, these consequences have been investigated experimentally \cite{UGMSFS10,KRBMGUIE12,NYHCKOLESUG10,NYHCKOLESUG11}.  Since the full counting statistics is performed by the capacitive coupling to a secondary circuit, back-action effects should be taken into account to interpret the experimental results.  Accordingly, the symmetry relation for the single
observed current holds with respect to an effective affinity which differs from the thermodynamic affinity
because of the noise coming from the secondary circuit \cite{UGMSFS10,BEG11}. In this context, new opportunities arise for a better understanding of the intriguing fact that quantum jumps can be observed continuously in time \cite{C93,GMWS01,GN99b}.  For electronic conduction in an external magnetic field, the current fluctuation theorem predicts relations beyond linear response between conductance and noise power \cite{SU08,AGMT09}.  Experimental observation on Aharonov-Bohm rings designed with GaAs semiconductor are in semiquantitative agreement with the theoretical predictions, possibly because of back action from the noise measurement \cite{NYHCKOLESUG10,NYHCKOLESUG11}.

In conclusion, the advent of time-reversal symmetry relations among the dynamical large-deviation properties has opened new perspectives in our fundamental understanding of nonequilibrium quantum systems.  These relations can be extended to many nonequilibrium systems beyond electron quantum transport.  In particular,  they may apply to fermion and boson quantum transport in the physics of cold atoms or molecules or in quantum optics. The properties which can be deduced from the symmetry relations notably concern the properties of energy transduction when several currents are coupled together in thermoelectric, photoelectric, or mass separation devices. In a broader perspective, time-reversal symmetry relations are also considered in relativistic and gravitational systems \cite{F07,IO11}.  Similar relations can be envisaged for other discrete symmetries or their combinations, e.g., in the context of quantum field theory \cite{CPT,CH08}. 

\begin{acknowledgments}
Useful discussions with Professors Markus B\"uttiker and Keiji Saito are kindly acknowledged.
This research has been financially supported by the F.R.S.-FNRS, 
the ``Communaut\'e fran\c caise de Belgique'' (contract ``Actions de Recherche Concert\'ees'' No. 04/09-312), and the Belgian Federal Government (IAP project ``NOSY").
\end{acknowledgments}



\begin{thebibliography}{99}
\bibitem{CPT} For high-energy systems involving all the forces of nature including the weak interaction, the corresponding symmetry is the combination of time reversal, space inversion, and charge conjugation.
\end{thebibliography}
\end{document}